**Profitability of Open-Source Software Product Development**

Shivendu P. Singh
Southern Illinois University
shising@siue.edu

Narayan Ramasubbu
University of Pittsburgh
narayanr@pitt.edu

Chris F. Kemerer
University of Pittsburgh
ckemerer@pitt.edu

M. Zia Hydari
University of Pittsburgh
zia.hydari@pitt.edu

***Abstract***

For-profit firms increasingly adopt open-source product development by engaging external community members alongside internal employees on social coding platforms such as GitHub. Yet whether and through what mechanisms this engagement affects firm profitability remains an open question. Drawing on the knowledge-based view of the firm, we conceptualize open-source product development as a form of distributed knowledge integration that enhances labor productivity by expanding the specialized expertise available for product development beyond the firm's internal boundaries. We posit that improvements in labor productivity translate into higher profitability, as labor constitutes a primary input in software product development. However, the labor productivity effect depends on the extent of participation by external contributors, and the resulting profitability gains are shaped by equifinal configurations of firm resource allocation. We examine our theoretical framework using a longitudinal dataset of 977 U.S. high-tech firms from 2001 to 2025 and find that firms adopting open-source product development via GitHub realized, on average, a 4%–5% increase in gross margin. These results are robust across staggered difference-in-differences, generalized synthetic control, instrumental variable, and dynamic panel specifications. A moderated mediation analysis decomposing over 323,000 project-level contributions across more than 44,000 repositories into internal employee and external volunteer sources reveals that labor productivity partially mediates the profitability effect and that this mediation is amplified by external contributor engagement. The indirect effect of open-source development intensity on profitability through labor productivity becomes discernibly positive only beyond a threshold of external volunteer contributions (approximately 35% in our sample). Configurational analysis further reveals that research and development intensity is present across all high-profitability configurations, consistent with the absorptive capacity required to integrate externally sourced knowledge. These findings extend the knowledge-based view to the open-source context and provide managerial guidance for aligning open-source strategies with firms' resource configurations.

## 1. Introduction

A vast majority of leading for-profit, high-tech firms maintain an active presence on the social coding platform GitHub, collectively hosting thousands of product development projects involving both employees and external community members (Peterson 2017; Haese and Peukert 2026). The shift from in-house, proprietary product development to more open, collaborative, and community-driven processes raises a fundamental question about whether firms can capture financial value from this strategy (Gambardella and von Hippel 2019; Wang et al. 2020; Mallipeddi et al. 2024).

Prior research has extensively explored many dimensions of open-source software development, including developer motivations (e.g., Shah 2006; Mehra et al. 2011), diversity of collectives (e.g., Daniel et al. 2013, 2018), firm adoption decisions (e.g., Nagle 2018), competitive dynamics (e.g., Zhu and Zhou 2012), licensing strategies (e.g., August et al. 2013), digital entrepreneurship (e.g., Lin and Maruping 2022), and community governance (e.g., Shaikh and Vaast 2016; O'Mahony and Karp 2022; Lindberg et al. 2024). However, whether open-source product development improves firm profitability remains empirically unestablished. As our thematic summaries in Online Appendix §A (Tables A1 and A2) reveal, the more than 50 studies we surveyed have predominantly relied on analytical modeling, qualitative case studies, or limited cross-sectional designs; none provides firm-level evidence on the profitability question using longitudinal data. While a few studies have modeled profitability gains from open-source engagement under certain competitive conditions (e.g., Kumar et al. 2011; Alexy et al. 2018; Mallipeddi et al. 2024), empirical examinations of the pathways from open-source engagement to firm profitability remain sparse.

Addressing this gap and drawing on the knowledge-based view of the firm (Kogut and Zander 1992; Grant 1996), we conceptualize open-source engagement during product development as distributed knowledge integration. When firms open their product development to external contributors on GitHub, they expand the pool of specialized expertise feeding into the development process. External contributors introduce diverse problem-solving approaches that accelerate internal work, and they supply situated technical knowledge drawn from heterogeneous deployment contexts, which the firm integrates with its internal architectural knowledge under its own governance. These collectively enhance a firm's labor productivity, which in turn drives gross margin improvement, given that labor is a key input in software development. We posit that not all firms' open-source engagement activates this mechanism equally. The labor productivity effect depends on whether external community members contribute a meaningful share of product development activities, a moderating condition that distinguishes genuine knowledge integration from the mere use of GitHub. The theoretical issue is therefore not whether firms can obtain external effort at lower cost, but whether and under what conditions firms can profitably integrate externally distributed knowledge into commercial product development. Moreover, firms differ in their resource allocations of research and development (R&D), capital, and labor investments and in the competitive dynamics of their markets, which may give rise to multiple configurations with equifinal profitability outcomes. We therefore propose a framework that encompasses the labor productivity channel, its moderation by the share of external contributions, and the resource configurations under which labor productivity gains translate into profitability.

To examine this framework empirically, we developed a longitudinal dataset of 977 U.S.-based high-tech firms from 2001 to 2025, combining GitHub engagement data with firm-level measures from Compustat and 10-K reports. In addition, we collected GitHub data on 323,634 project-level events across 44,147 repositories and used a multi-stage heuristic that separates internal employee and external volunteer activity. Our findings indicate that firms adopting open-source product development via GitHub experienced, on average, a 4%–5% increase in gross margin relative to non-adopting peers. This result is robust across a variety of specifications involving staggered difference-in-differences (DID) analysis (Callaway and Sant'Anna 2021), the de Chaisemartin and D'Haultfœuille (2026) DID estimator, generalized synthetic control (Xu 2017), instrumental variable regression, and dynamic panel models (Arellano and Bond 1991; Blundell and Bond 1998). A moderated mediation analysis, estimated using generalized structural equation modeling, confirms that labor productivity partially mediates the profitability effect and that external contributor engagement amplifies this mediation. The indirect effect of open-source development intensity on profitability through labor productivity becomes discernibly positive only when volunteer contributions exceed a threshold of total observed development activity (approximately 35% in our sample). Configurational analysis using fuzzy-set qualitative comparative analysis (fsQCA) further reveals that R&D intensity is present across all high-profitability configurations, consistent with our expectation that firms require sufficient internal knowledge to recognize, assimilate, and exploit externally sourced contributions.

## 2. Open-Source Product Development: Background and Conceptual Development

### *2.1 Open-Source Product Development as Distributed Knowledge Integration*

The knowledge-based view of the firm posits that competitive advantage derives from a firm's ability to integrate specialized knowledge held by individuals (Kogut and Zander 1992; Grant 1996). Recent work continues to treat knowledge as a central strategic resource while shifting attention to the organizational routines and boundary-spanning mechanisms through which distributed knowledge becomes integrated and valuable (Grant and Phene 2022; Bergh et al. 2025). A parallel evolution in the absorptive-capacity literature moves from the recognition and application of external knowledge (Cohen and Levinthal 1990) toward routines for acquiring, assimilating, transforming, and exploiting it (Zahra and George 2002; Todorova and Durisin 2007), including in digital and online-community settings where knowledge is accessed and integrated across organizational boundaries (Huang et al. 2022; Gol et al. 2024). This contemporary view is well suited to open-source product development, where a firm must not only reach external contributors but also evaluate, coordinate, and integrate their contributions into its development processes and product architecture.

Traditional modes through which software firms acquire external knowledge, such as hiring, consulting, or strategic alliances, involve significant search costs, contractual overhead, and integration frictions (Cassiman and Valentini 2016). Open-source product development offers a fundamentally different approach to expanding the accessible knowledge base. When a for-profit firm adopts open-source product development via platforms such as GitHub, it opens its development processes to contributions from a distributed community of external developers. These external contributors, who may include independent developers, employees of partner or customer firms, and academic researchers, bring specialized technical expertise, problem-solving approaches, and domain-specific knowledge drawn from diverse organizational contexts (von Hippel and von Krogh 2003; Setia et al. 2012). The firm's internal employees work within an expanded production function that incorporates external knowledge inputs, effectively broadening the knowledge base available for product development without a commensurate increase in labor costs.

### *2.2 From Open-Source Engagement to Labor Productivity*

Open-source product development enhances labor productivity through the integration of externally distributed knowledge with the firm's internal expertise, which operates through at least two reinforcing dynamics: mobilizing situated external expertise and absorbing diverse problem-solving knowledge. External contributions to open-source projects extend well beyond routine tasks and carry substantive knowledge content. Even a bug fix or a code review can reveal an edge case, an unanticipated deployment environment, an integration failure, or a violated architectural assumption, which are all difficult to observe internally. Community pull requests, issue discussions, and reviews encode technical judgment drawn from heterogeneous deployment contexts (Faraj et al. 2011; Xiao et al. 2018; Shaikh and Vaast 2023). Open-source communities shape the novelty and complexity of the software itself, not merely its maintenance (Lindberg et al. 2024), and firms increasingly build important, and in some cases core, commercial products on open-source foundations rather than opening only non-differentiating utilities (Conti et al. 2025; Haese and Peukert 2026). Engaging external contributors thus mobilizes situated expertise drawn from heterogeneous use contexts, technology stacks, and user needs that the focal firm cannot readily observe internally (Setia et al. 2012; Riehle et al. 2014; Gambardella et al. 2017). To capture these knowledge inputs, firms structure open-source projects to coordinate internal and external contributions while retaining architectural governance (West and O'Mahony 2008; Di Tullio and Staples 2013). By incorporating external contributors into the development process, the firm effectively broadens the knowledge base feeding into product development without a proportional increase in internal headcount. Furthermore, open-source engagement exposes internal employees to diverse knowledge inputs and novel problem-solving approaches (Ågerfalk and Fitzgerald 2008; Faraj et al. 2016). Singh et al. (2011) documented learning dynamics within open-source communities whereby developers accumulate knowledge through interaction with heterogeneous peers, and Tang et al. (2020) showed that access to community knowledge yields productivity gains only to the extent that a project has the internal knowledge base to absorb it, a point we develop as a configurational boundary condition (§2.5). When firms possess this capacity, knowledge spillovers from community engagement enhance internal problem-solving and accelerate development cycles (Huang et al. 2022; Gol et al. 2024). Together, these dynamics provide the conceptual basis for expecting that firms adopting open-source product development will experience improvements in the productivity of their internal workforce.

### *2.3 From Labor Productivity to Profitability*

The link between labor productivity and profitability is grounded in the economics of software development. Profitability reflects the proportion of revenue retained after deducting direct production and delivery costs. In software firms, labor constitutes the dominant component of direct costs, as the primary inputs to product development are the knowledge and skills of the workforce rather than raw materials or physical capital. When labor productivity increases, meaning the firm generates more revenue per employee, profitability improves through two complementary routes. First, if productivity gains manifest as greater output from the same workforce (e.g., higher-quality products or faster releases), revenue increases while direct costs remain stable. Second, if productivity gains manifest as equivalent output with reduced internal labor commitment, because integrating external contributions lowers the internal effort required for a given output, direct costs decrease while revenue holds. In practice, both routes likely operate simultaneously, making labor productivity a structurally proximate driver of firm gross margin. We note that this mediation is expected to be partial rather than complete. Open-source engagement may influence profitability through additional mechanisms beyond labor productivity, such as community governance and responsive engagement (Di Tullio and Staples 2013; Germonprez et al. 2017), licensing-driven value generation (August et al. 2018), or signaling benefits and access to external knowledge that facilitates innovation and funding (Conti et al. 2025). While these alternative pathways are beyond the scope of the present study, we acknowledge their potential contribution to any remaining direct effect.

### *2.4 The Role of External Contributors*

The labor productivity gains we theorize are not expected to accrue uniformly across all firms that simply maintain a presence on GitHub. The theorized labor productivity effect depends critically on the degree to which external contributors are actually involved in the firm's product development process, shaped by the participation architecture that governs community access (West and O'Mahony 2008; Di Tullio and Staples 2013; Zhang et al. 2026). To capture this distinction conceptually, we differentiate between internal (employee) and external (volunteer) contributions to a firm's open-source repositories (cf. Mockus et al. 2002; Zhang et al. 2026). The relative share of external contributions is an observable indicator of the extent to which the external knowledge community is meaningfully activated in the firm's product development processes: a higher share indicates that the firm has successfully attracted substantive community participation and is drawing on distributed problem solving and situated expertise, not merely hosting repositories on a social coding platform. We therefore expect the relative share of external contributions to moderate the pathway from open-source development intensity to labor productivity, such that the positive productivity effect is amplified when external contributors constitute a meaningful share of the project workforce. This moderated mediation formulation allows us to move beyond the question of whether open-source engagement improves productivity to specify under what conditions of community participation it does so.

### *2.5 Resource Configurations*

The translation of labor productivity gains into profitability improvements is unlikely to be uniform across firms. Drawing on the concept of absorptive capacity (Cohen and Levinthal 1990) and the literature on resource complementarities (Milgrom and Roberts 1995; Ennen and Richter 2010), we expect that the effectiveness of the knowledge integration mechanism depends on the firm's internal resource configuration and its competitive environment. Beyond the two focal constructs, open-source development and the share of external contributions, we consider six configurational dimensions that collectively shape this translation: (1) R&D, which enables the firm's absorptive capacity for external knowledge; (2) capital, which provides the infrastructure for integrating contributions; (3) labor, which reflects the internal workforce available to orchestrate and integrate external inputs; (4) firm size, which conditions the scale and governance of community engagement; (5) market share, which accounts for the firm's competitive position; and (6) industry concentration, which reflects the competitive structure of the firm's specific industry sector that shapes the incentives to draw on external and distributed knowledge sources.

Cohen and Levinthal (1990) established that a firm's ability to recognize, assimilate, and exploit external knowledge depends on its prior stock of related knowledge, which is substantially determined by R&D investments. Subsequent work reconceptualizes this capacity as a set of routines for acquiring, assimilating, transforming, and exploiting external knowledge (Zahra and George 2002; Todorova and Durisin 2007). In the context of open-source product development, R&D investments reflect the firm's capacity to evaluate external contributions, integrate them into its product architecture, and extend them into commercially viable features. Firms with low R&D may lack the technical depth to filter and absorb community contributions effectively, leading to unproductive coordination costs that offset profits. A firm's financial capital is required to build the technological infrastructure, such as continuous integration and deployment pipelines, automated testing frameworks, and code review systems, which facilitate the efficient incorporation of external contributions into the firm's development workflow (Shaikh and Vaast 2023). A firm's available labor pool reflects the depth of the firm's internal workforce, which serves as the integration core that orchestrates external contributions, maintains architectural coherence, and ensures alignment between open-source outputs and commercial objectives (Di Tullio and Staples 2013; Shaikh and Henfridsson 2017). Firm size may further condition these dynamics, as larger firms typically command greater resources for community governance but may also face bureaucratic inertia that slows the absorption of externally sourced knowledge (Harison and Koski 2010; Bogers et al. 2018). Finally, the firm's market position and the competitive structure of the firm's industry, as captured by market share and industry concentration, shape both the incentive and the ability to extract value from open-source engagement. Firms in more competitive, less concentrated markets may face stronger pressure to leverage distributed knowledge sources as a basis for differentiation, while firms in concentrated industries may have less impetus to invest in community-driven development (Boudreau 2010).

Rather than modeling these factors as independent moderators, we adopt a configurational perspective that recognizes their joint and interdependent influence on the open-source profitability effect (cf. Ramasubbu and Bardhan 2021; Mithas et al. 2022). This perspective accommodates two properties especially relevant to our context: equifinality, which is the possibility that different resource configurations may yield equivalent profitability outcomes, and causal asymmetry, which is the possibility that the conditions enabling high profitability may differ structurally from those associated with low profitability. By examining how open-source product development, external contributor engagement, R&D, capital, labor, firm size, market share, and industry concentration combine to produce high or low profitability outcomes among open-source adopters, the configurational approach reveals the specific resource profiles under which the knowledge integration mechanism translates into profitability.

## 3. Data and Empirical Approach

### *3.1 Data Collection*

We focused on the U.S. high-tech industry, a vital sector accounting for approximately 19% of all private sector employment and 23% of the country's total industrial output (Wolf and Terrell 2016; Atkinson 2022). To define this sector, we followed the industry coding schemes used in prior studies published by the Bureau of Labor Statistics (BLS) and the U.S. Census Bureau (Wolf and Terrell 2016; Goldschlag and Miranda 2020; Atkinson 2022). This definition includes firms in sectors such as software publishers, data processing, telecommunications, professional services, and electronic products, which typically employ at least five times the national average share of STEM workers (Goldschlag and Miranda 2020). The complete list of industry codes is provided in Online Appendix §B. We selected firms from those NAICS codes that had available operating and financial performance data in the Compustat database, resulting in an initial sample of 1,746 firms. After thorough data cleaning to filter out missing values, erroneous and duplicate entries, foreign firms listed as American Depositary Receipts (ADRs), firms that do not list revenues in USD, and single-entry firms, we retained 1,460 high-tech firms. We then determined each firm's involvement in open-source product development through a comprehensive search across major open-source community platforms, including GitHub, GitLab, SourceForge, and Bitbucket. Additional details about our search procedures and GitHub data extraction are presented in Online Appendix §B. Out of the 1,460 firms, we identified 305 firms as engaged in open-source product development for at least one year between 2011 and 2025; all of these firms maintained a presence on GitHub. Following recommended practice for quasi-experimental designs, we removed 8 always-adopted firms (i.e., firms with no pre-adoption data) for clean comparative analysis (Callaway and Sant'Anna 2021).

In the next step, we matched our data with the Compustat database to obtain firm-level financial and operational metrics. We had initially identified 305 firms engaged in open-source product development through GitHub. After excluding the 8 always adopters and 66 firms lacking sufficient Compustat data in either the pre- or post-adoption periods, we retained 231 adopting firms. Overall, the matching yielded a final sample of 977 U.S.-based high-tech firms (up to 11,854 total firm-year records across empirical specifications). The resulting dataset forms an unbalanced panel with staggered GitHub adoption, comprising 231 adopters and 746 non-adopters. Of the 977 firms, 56 have data for all 25 years (2001–2025), with 23 adopting GitHub between 2011 and 2025. The remaining 921 firms have data for fewer than 25 years; of these, 208 hosted projects on GitHub between 2011 and 2025. This group exhibits various data patterns: 38 firms are left-censored, 51 firms are right-censored, and 119 have a combination of censorship and missing-at-random issues. For treated firms, the median tenure on GitHub was 7.4 years (min. 1, max. 15). The unbalanced panel structure and staggered adoption patterns are visually represented in Figure C1 of Online Appendix §C, and Figure C2 shows profitability distribution for adopters and non-adopter firms.

To assess the share of external contributions to a firm's open-source repositories, we collected firm-level engagement data using the GitHub API and GitHub Archive (through Google BigQuery). We extracted detailed data, including the number of projects, project registration dates, and firm registration dates. In addition, we collected 323,634 events (e.g., Push, PullRequest, IssueEvent, ForkEvent, and PullRequestReviewComment) for 44,147 repositories hosted by all the 231 firms that adopted GitHub for product development in our final sample. To classify contributions as internal or external, we identified unique contributors through a combination of their public actor.login and the author.email addresses extracted from the GitHub archives. We applied a multi-stage filtration heuristic that classifies as "internal" any account associated with the firm's verified corporate email domains or inferred from a combination of behavioral signals, including work-hour activity patterns and forms of participation in governance activities such as pull-request review (Mockus et al. 2002; Riehle et al. 2014; Zhang et al. 2026). The remaining distinct identities are classified as external contributors. We aggregated these classifications at the firm-year level to measure external and internal participation by year.

### *3.2 Variables*

***Dependent Variable***. *Gross Margin*, defined as log of *(Total Sales − Cost of Goods Sold) / Total Sales*, is a widely used profitability measure for assessing a firm's financial performance (e.g., Zhu and Kraemer 2002; Ederhof et al. 2021; Stamatopoulos et al. 2021). *Gross Margin* is particularly suited to our context because it reflects the financial value generated by product development processes after accounting for direct costs (i.e., production-level profitability). As discussed in §2.3, because labor constitutes the main input in software product development, *Gross Margin* is structurally sensitive to changes in labor productivity, the mediating mechanism that we theorize.

***Independent Variables***. We employ two measures of open-source engagement. The binary variable, *Open-Source Development*, indicates whether a firm was engaged in open-source product development on GitHub in a given year. The continuous variable, *Open-Source Development Intensity* (*OSDI*), reflects the depth of this engagement, measured as the natural logarithm of the total number of active projects of a firm in a year. An active project is identified as one having code contributions including commits and issue resolutions (extracted from PushEvent, PullRequestEvent, PullRequestReviewCommentEvent, and IssueEvent) in a given year by any employee or external volunteer.

***Mediating Variable.*** *Labor Productivity* is operationalized as total sales per employee (log-transformed). As detailed in §2.3, this measure captures the aggregate efficiency with which a firm converts its workforce into revenue-generating output. Revenue per employee is widely recognized as a valid proxy for human capital efficiency in both academic research (e.g., Zhu and Kraemer 2002; Hancock et al. 2013; Andrade-Rojas et al. 2024) and professional practice, where it serves as a key benchmark for organizational productivity and resource allocation decisions (Wiggins 2022; APQC 2024). We selected this measure because it is standardized, output-oriented, and externally comparable across firms.

***Moderating Variable.*** The *Volunteer Contribution Ratio* (*VCR*) is the number of unique observed external (non-employee) contributions divided by total observed contributions (external + internal) across a firm's open-source repositories each year. Unique contributions combine multiple event types, including commits, pull requests, issues, and reviews, to identify distinct interactions with a firm's repositories from external volunteers. Each distinct event (commit, pull request, issue, or review) constitutes one contribution. This ratio-based measure offers several advantages over alternatives such as raw counts of external commits or contributor headcounts. First, it captures the relative structure of a firm's internal–external labor mix, reflecting its orientation toward openness (West and O'Mahony 2008). Second, it mitigates size-related confounds, as absolute commit volumes strongly correlate with firm size, repository count, and project maturity. Third, prior software-engineering research shows that relative measures of external participation facilitate comparisons across firms and over time, whereas raw contribution volumes conflate scale effects with participation structure (Mockus et al. 2002; Riehle et al. 2014; Zhang et al. 2026). For robustness checks, we used a narrower, alternative measure, *Volunteer Pull Request Ratio (VPRR)*, which only accounts for code-related contributions. More details about this measure and related analysis are reported in Online Appendix §F.

***Control Variables.*** Building on prior studies (e.g., Brynjolfsson and Hitt 2003; Nagle 2019), we include several controls that are believed to influence profitability: *R&D* (log of research and development investments calculated using the perpetual inventory method with 15% depreciation), *Capital* (log of net capital with 5% depreciation), *Employees* (log of employee count), *Large Firm* (binary indicator for firms in the top 25$^{th}$ percentile by industry sales), *Market Share* (firm sales relative to NAICS-2 sector sales), and *Industry Concentration* (Herfindahl-Hirschman Index). In the moderated mediation models, we additionally control for aggregate project-level variables, *Total Forks*, *Total Stars*, and *Total Issues* raised per firm-year, to account for project popularity and quality characteristics that may independently affect *Labor Productivity*. These firm-resource and market variables serve as controls in the econometric analyses; in the configurational analysis (§3.3 and §4.3), a subset is reinterpreted as configurational conditions. In that analysis, *Labor* denotes the firm's internal labor input, normalized by firm revenue, and is the revenue-scaled counterpart of the *Employees* control; *R&D* and *Capital* are likewise expressed as revenue-normalized intensities.

All variables are tabulated with definitions and data sources in Table C1 of Online Appendix §C.

### *3.3 Modeling Approach*

We deploy a three-stage empirical approach that progressively deepens the analysis from establishing the average causal effect, to uncovering the mechanism, to identifying the boundary conditions corresponding to the theoretical framework developed in §2.

***Stage-1: Average Causal Effect.*** We first estimate the average firm-level profitability gains associated with GitHub adoption using multiple program evaluation methods. Our primary specification is the staggered difference-in-differences (DID) approach proposed by Callaway and Sant'Anna (2021), which accommodates multiple treatment periods, variation in treatment timing, and our unbalanced panel structure. In addition, to establish robustness, we employ the DID estimator proposed by de Chaisemartin and D'Haultfœuille (2026), which accommodates intertemporal treatment reversals, and the generalized synthetic control method (Xu 2017), which relaxes the parallel trends assumption. We further conduct an instrumental variable analysis using industry-level peer adoption of GitHub as the instrument (cf. Bendig et al. 2023; Gao et al. 2026) to address endogeneity concerns. Beyond the binary adoption measure, we estimate the marginal effect of *OSDI* on *Gross Margin* using dynamic panel models, the Arellano-Bond (ABOND) and Blundell-Bond (BBOND) estimators (Arellano and Bond 1991; Blundell and Bond 1998), which use lagged instruments to address potential issues related to endogeneity and unobserved heterogeneity.

***Stage-2: Moderated Mediation Mechanism.*** For investigating the labor productivity mechanism theorized in §2.3–§2.4, we employ Generalized Structural Equation Modeling (GSEM) to simultaneously estimate the direct effect of *OSDI* on *Gross Margin* and the indirect effect via *Labor Productivity*, conditional on *VCR*. Specifically, we estimate a moderated mediation framework corresponding to Model 7 as outlined by Hayes (2017). In this specification, *VCR* moderates the path from *OSDI* to *Labor Productivity* (the first stage of mediation), while the path from *Labor Productivity* to *Gross Margin* and the direct path from *OSDI* to *Gross Margin* are estimated simultaneously.

***Stage-3: Configurational Analysis.*** To explore the heterogeneity in the open-source profitability effect, we employ fuzzy-set Qualitative Comparative Analysis (fsQCA). This approach identifies the specific configurations of firm-level factors that consistently produce high or low profitability among the GitHub adopters, while accommodating equifinality and causal asymmetry (Ramasubbu and Bardhan 2021; Mithas et al. 2022). The configurational conditions include *OSDI*, *VCR*, *Capital*, *Labor*, *R&D*, *Large Firm*, *Market Share,* and *Industry Concentration*. We normalize *R&D*, *Capital*, and *Labor* using firm revenues to derive intensity measures, collapse the panel structure into three-year post-adoption averages, and calibrate all variables into fuzzy-set membership scores ranging from 0 to 1.

This three-stage approach ensures that each analytical step serves a clear function: Stage-1 establishes whether open-source engagement improves profitability and demonstrates robustness; Stage-2 uncovers how it does so and identifies the critical role of external contributor participation; and Stage-3 reveals under what firm-level resource configurations the profitability gains are best realized.

## 4. Analysis and Results

### *4.1 Stage-1: Average Causal Effect of Open-Source Product Development on Profitability*

Table 1 presents the results of our primary analysis using the Callaway and Sant'Anna (2021) staggered DID estimator. Because Microsoft Corp. acquired GitHub in 2018, potentially affecting adoption and use patterns, we present two sets of results: Model 1 restricts the sample to the pre-acquisition period, and Model 2 uses the full sample through 2025. The coefficient of *Open-Source Development* is positive and significant in both specifications. The results indicate that, all else being equal, firms in our sample increased their gross margins by approximately 4% on average by adopting open-source product development. The average treatment effect on the treated (ATT) aggregated by group and by calendar year shows consistently positive effects. The tests of conditional parallel trends were not significant (pre-acquisition sample: $\chi^2 = 50.64$, $p= 0.99$; full sample: $\chi^2 = 28.35$, $p=1.00$), indicating no evidence of violations of this critical assumption. ATT across relative adoption years (−10 to 14) is shown in Figure D1 of Online Appendix §D, confirming that the cumulative effect is not significantly different from zero in the pre-adoption period but steadily increases during the post-adoption period. Winsorized results are consistent and reported in Online Appendix Table D1. To further test the robustness of our findings, we estimated the effect of GitHub adoption on *Gross Margin* using several alternative specifications, including the DID estimator proposed by de Chaisemartin and D'Haultfœuille (2026), generalized synthetic control method (Xu 2017), and instrumental variable analysis using industry peer adoption at the 2-digit NAICS and SIC levels as instruments (cf. Bendig et al. 2023; Gao et al. 2026). These additional results are presented in Online Appendix §D; all robustness estimates yield positive and significant effects, broadly consistent with the 4%–5% baseline estimate.

Going beyond the binary adoption measure, we estimate the marginal effect of *OSDI* on *Gross Margin* using dynamic panel methods. Table 2 presents results from three specifications: two-way fixed effects (TWFE, Model 1), Arellano-Bond (ABOND, Model 2), and Blundell-Bond (BBOND, Model 3). All three consistently reveal a positive effect of development intensity on profitability, with an estimated elasticity near 0.01 (i.e., a 10% increase in a firm's open-source projects is associated with approximately a 0.1% increase in *Gross Margin*). Results for data prior to Microsoft's acquisition of GitHub and Winsorized results are consistent and reported in Online Appendix §E. Taken together, the Stage-1 results establish that open-source software development has a positive and robust effect on firm profitability, whether measured through binary adoption or continuous development intensity, and whether estimated via staggered DID, generalized synthetic control, instrumental variables, or dynamic panel methods.

**Table 1. Open-Source Product Development and Profitability**
Dependent Variable = *Gross Margin*

| Independent Variable | Staggered DID using data prior to Microsoft's acquisition of GitHub *Model 1* | Staggered DID using the full sample *Model 2* |
|---|---|---|
| Open-Source Development (ATT) | .044*** (.015) | .044*** (.016) |
| ATT by Group | .042*** (.014) | .040*** (.015) |
| ATT by Calendar Year | .044*** (.016) | .041** (.017) |
| Control Variables† | Yes | Yes |
| Firm Fixed Effect | Yes | Yes |
| Industry Fixed Effect | Yes | Yes |
| Time Fixed Effect | Yes | Yes |
| Observations | 8,878 | 11,797 |
| Total No. of firms | 973 | 977 |
| No. of treated firms | 197 | 231 |
| Parallel-trend test | 50.64 *(p=0.99)* | 28.35 *(p =1.00)* |

*Note:* Standard errors are in parentheses*; *** p<.01, ** p<.05, * p<.1;* Control variables included are *R&D*, *Capital*, *Employees*, *Large Firm*, *Market Share,* and *Industry Concentration*. The *csdid* estimator (Callaway and Sant'Anna, 2021) utilizes control variables to adjust for conditional parallel trends via inverse probability weighting and/or outcome regression. Consequently, it does not produce or report standalone linear coefficients for these covariates.

**Table 2. Open-Source Development Intensity and Profitability** *(full sample)*
Dependent Variable = *Gross Margin*

| Independent Variable | TWFE (1) | ABOND (2) | BBOND (3) |
|---|---|---|---|
| Open-Source Development Intensity | .010*** (.004) | .009*** (.002) | .012*** (.002) |
| Control Variables† | Yes | Yes | Yes |
| Firm Fixed Effect | Yes | Yes | Yes |
| Industry Fixed Effect | Yes | Yes | Yes |
| Time Fixed Effect | Yes | Yes | Yes |
| Observations | 11,854 | 10,095 | 11,154 |
| Total No. of firms | 977 | 908 | 957 |

*Note:* Standard errors are in parentheses*; *** p<.01, ** p<.05, * p<.1;* the number of observations in models (2) and (3) differ from those in model (1) due to the use of lagged instrumental variables and unavailability of data for certain lagged calendar years*;* Control variables included are *R&D*, *Capital*, *Employees*, *Large Firm*, *Market Share,* and *Industry Concentration*.

### *4.2 Stage-2: Moderated Mediation Mechanism*

To investigate the labor productivity mechanism, we conducted moderated mediation analysis using adopter firms' data. Using GSEM, we simultaneously estimated the direct effect of *OSDI* on *Gross Margin* and the indirect effect via *Labor Productivity*, conditional on *VCR*. Moderated mediation effects were assessed using 5,000-replication bootstrapping. We compared a base mediation model (Model 1, without the OSDI × VCR interaction) against the full moderated mediation model (Model 2, with the interaction term). Table 3 reports the path coefficients for both models. The full model yields lower values for both AIC (−316.285 vs. −310.785) and BIC (−44.072 vs. −44.016) relative to the base model, indicating that the inclusion of the interaction term provides a superior fit even after penalizing for additional model complexity.

**Table 3. Open-Source Development Intensity, Labor Productivity, and Profitability**

| **DV** | **Predictor** | **Model 1 (Base)** | **Model 2 (Full)** |
|---|---|---|---|
| Labor Productivity | Constant | 2.685***(.154) | 2.694***(.154) |
| | Open-Source Development Intensity (*OSDI*) | .013**(.006) | .011* (.006) |
| | Volunteer Contribution Ratio (*VCR*) | .103** (.040) | -.003 (.056) |
| | *OSDI* X *VCR* | | .069***(.025) |
| | Control Variables† | Yes | Yes |
| | Year Fixed Effect | Yes | Yes |
| Gross Margin | Constant | -.305*** (.089) | -.305***(.089) |
| | Labor Productivity | .196***(.013) | .196***(.013) |
| | Open-Source Development Intensity (*OSDI*) | .011***(.003) | .011***(.003) |
| | Year Fixed Effect | Yes | Yes |
| | Control Variables† | Yes | Yes |
| Number of firm years | | 1710 | 1710 |
| Number of firms | | 231 | 231 |
| Pseudo Log-Likelihood | | 204.392 | 208.142 |
| Akaike Information Criterion (AIC) | | -310.785 | -316.285 |
| Bayesian Information Criterion (BIC) | | -44.016 | -44.072 |

*Note:* Standard errors are in parentheses*; *** p<.01, ** p<.05, * p<.1;* Control variables included are *R&D*, *Capital, Employees*, *Large Firm*, *Market Share,* and *Industry Concentration, Total Forks, Total Stars, Total Issues raised by firm per year.*

In the full interaction model, *VCR*'s direct effect (β = −0.003, N.S.) on *Labor Productivity* becomes non-significant and *OSDI*'s is reduced to marginal significance (β = 0.011, p<0.1), whereas both are significant in the base model (OSDI: β = 0.013, p < 0.05; VCR: β = 0.103, p < 0.05). The interaction between *OSDI* and *VCR* on *Labor Productivity* is positive and significant (β = 0.069, p < 0.01), confirming that external volunteer engagement amplifies the effect of open-source development intensity on labor productivity. This supports the theoretical expectation developed in §2.4: the distributed knowledge integration is activated to the extent that external contributors constitute a meaningful share of the development workforce. This finding confirms that the labor productivity mechanism depends on firms effectively mobilizing and integrating external knowledge through community participation, not merely maintaining a GitHub presence. The path from *Labor Productivity* to *Gross Margin* is positive and significant (β = 0.196, p < 0.01), confirming the structural link theorized in §2.3: because labor constitutes the key input to software product development, improvements in labor productivity translate into gross margin gains. Finally, the direct path from *OSDI* to *Gross Margin* remains significant (β = 0.011, p < 0.01), indicating partial mediation. This is consistent with our theoretical expectation (§2.3) that open-source engagement may influence profitability through additional mechanisms beyond labor productivity. These results are summarized in Figure 1.

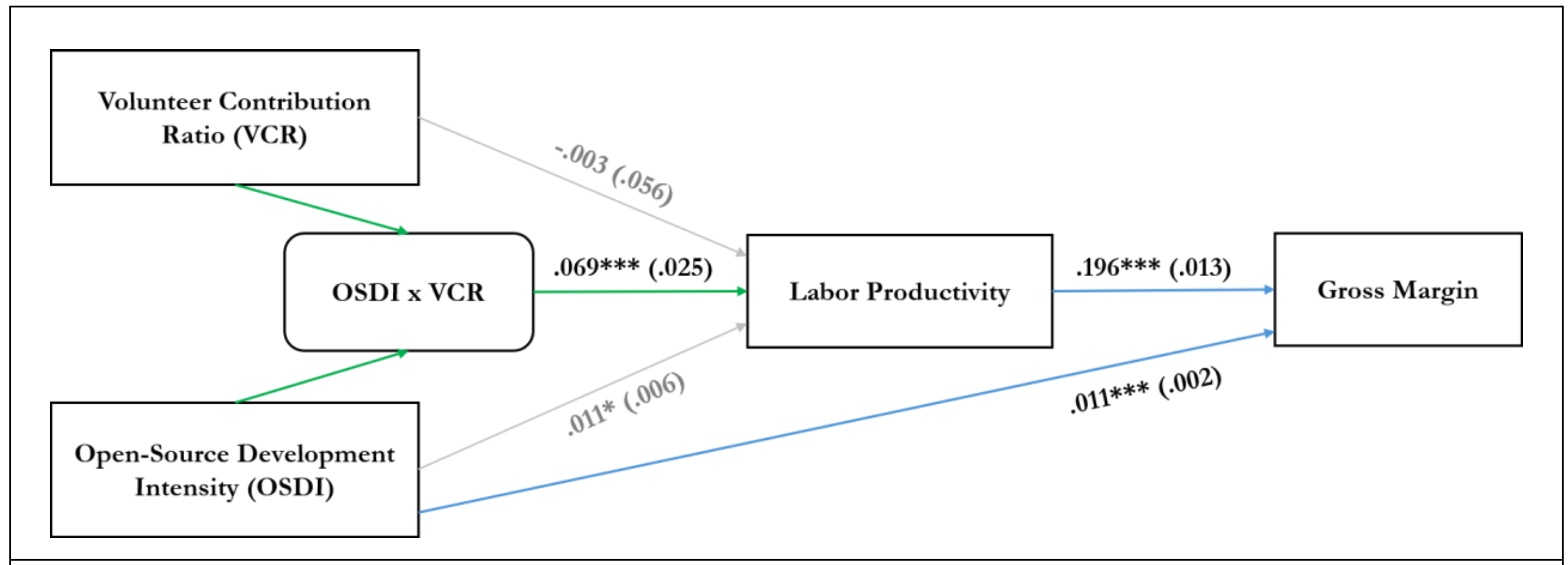

**Figure 1: Moderated Mediation Path Coefficients**
*Note:* *** p < .01; standard errors in parentheses. Gray paths are main effects in the full model (*OSDI* β=.011(.006), p<.1; *VCR* β=-.003(.056), N.S.). The green path denotes the first-stage interaction.

Table 4 reports the bootstrapped conditional indirect effects at different levels of *VCR*. The indirect effect of *OSDI* on *Gross Margin* via *Labor Productivity* is non-significant at low levels of external contribution (p>0.1), not conventionally significant at the mean level (p = 0.052), and strongly significant at high levels (p < 0.01). The Index of Moderated Mediation is significant (ω = 0.014, p<0.01), confirming that the strength of the mediation is statistically dependent on the level of external volunteer engagement. Results for data prior to Microsoft's acquisition of GitHub and Winsorized results are consistent and reported in Online Appendix §F. We further estimated the moderated mediation model using the narrower measure of volunteer contribution based on the *Volunteer Pull Request Ratio (VPRR)*; the results are consistent with our main findings and reported in Online Appendix §F.

Figure 2 visualizes the conditional indirect effect across the full range of *VCR* values, showing that the mediated effect becomes discernibly positive at approximately 35% external contribution. This threshold provides a practically interpretable benchmark: firms in our sample whose open-source projects attract at least 35% of their development activity from external volunteers realize significant profitability gains through the labor productivity mechanism, while those below this threshold do not. We interpret this as a sample-specific point at which the labor-productivity pathway becomes positive and statistically discernible, not as a universal managerial cutoff. To summarize Stage-2 results, the moderated mediation analysis provides evidence that labor productivity partially mediates the relationship between open-source development intensity and profitability, and that this mediation is conditional on the degree of external contributor engagement, which are consistent with the knowledge integration framework developed in §2.

**Table 4. Conditional Indirect Effects at Different Levels of Volunteer Contribution Ratio**

| Levels of *VCR* | Indirect Effect | SE |
|---|---|---|
| Low (-1 SD) | -.001 | .002 |
| Mean | .002* | .001 |
| High (+1 SD) | .005*** | .001 |
| Index of Moderated Mediation | .014*** | .005 |

*Note: *** p<.01, ** p<.05, * p<.1*

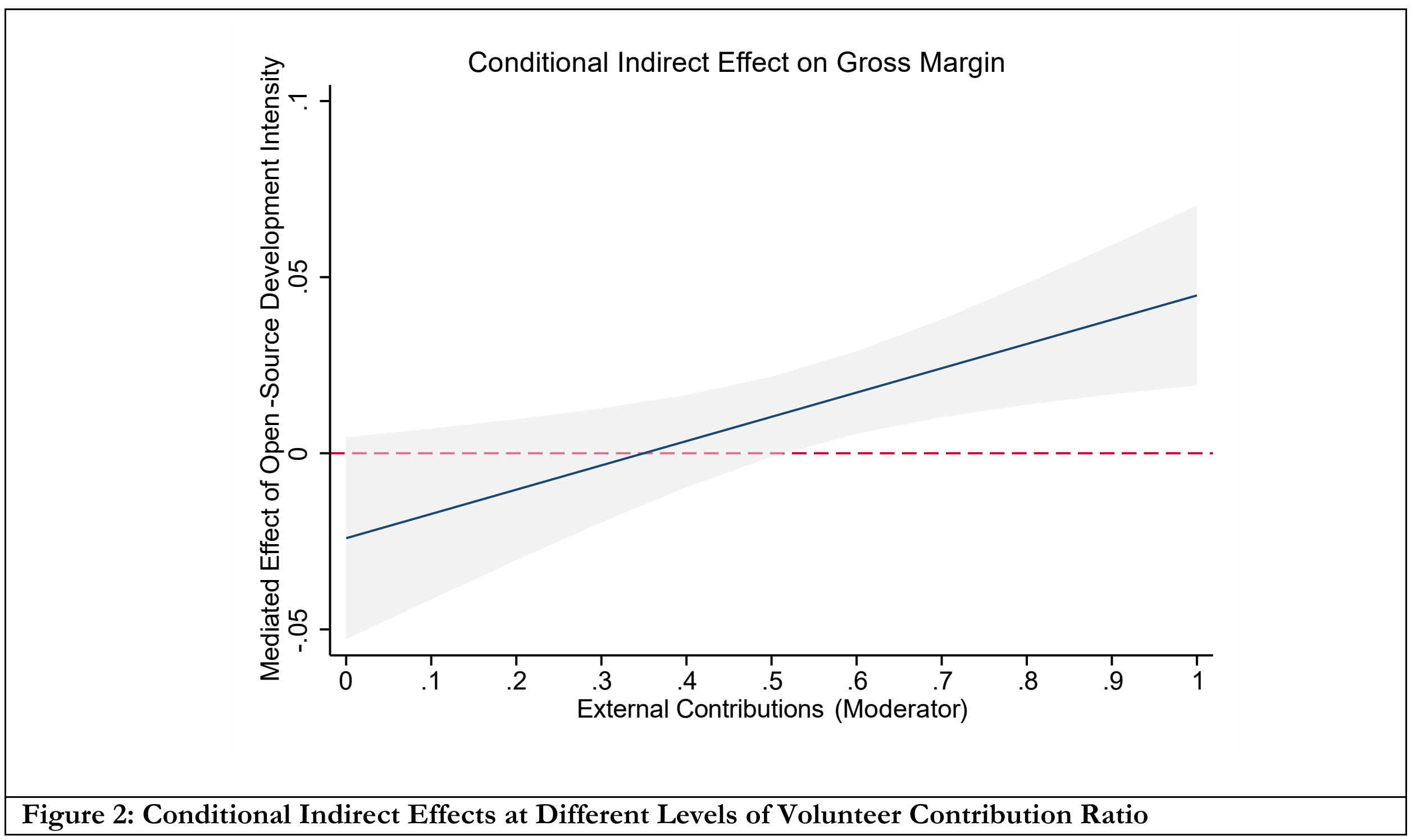

**Figure 2: Conditional Indirect Effects at Different Levels of Volunteer Contribution Ratio**

### *4.3 Stage-3: Configurational Analysis*

We employ fsQCA to identify the specific configurations of firm-level factors that consistently produce high or low profitability among open-source adopters. We normalized *R&D*, *Capital*, and *Labor* by firm revenue and calculated three-year averages of these measures post-adoption. Following standard practice, we calibrated all variables into fuzzy-set membership scores ranging from 0 to 1. Table 5 presents the results. The analysis reveals five configurations that consistently produce high profitability and four configurations that consistently yield low profitability following the adoption of open-source product development.

**Table 5. Resource Configurations**

| | **Open-Source Software Product Development Firms** | | | | | | | | |
|---|---|---|---|---|---|---|---|---|---|
| | High Profitability Configurations | | | | | Low Profitability Configurations | | | |
| **Variables** | 1 | 2 | 3 | 4 | 5 | 1 | 2 | 3 | 4 |
| OSDI | ● | ● | ● | ● | ● | ⊗ | ⊗ | ⊗ | ⊗ |
| VCR | ⊗ | ● | ⊗ | ● | ⊗ | ● | ⊗ | ⊗ | ● |
| Capital | ⊗ | ⊗ | ⊗ | ⊗ | ● | ● | ● | ● | ● |
| Labor | ● | ● | ● | ⊗ | ● | ● | ● | ● | ● |
| R&D | ● | ● | ● | ● | ● | ⊗ | ⊗ | ● | ● |
| Large Firm | ● | ● | ⊗ | ● | ● | ● | ● | ● | ● |
| Market Share | ⊗ | ⊗ | ⊗ | ● | ● | ● | ● | ● | ● |
| Industry Concentration | ⊗ | ⊗ | ⊗ | ⊗ | ⊗ | ● | ● | ● | ● |
| **Consistency** | 0.979 | 0.978 | 0.977 | 0.976 | 0.976 | 0.962 | 0.960 | 0.949 | 0.944 |
| **Raw Coverage** | 0.286 | 0.288 | 0.353 | 0.287 | 0.287 | 0.168 | 0.166 | 0.192 | 0.195 |

*Note: ● High level of the variable; ⊗ Low level of the variable.*

The configurations reveal several patterns that align with the theoretical framework developed in §2.5. First, *OSDI* and *R&D* are present in all five high-profitability configurations, which is consistent with the absorptive capacity prediction that firms require sufficient *R&D* investment, reflecting the internal knowledge stock necessary to recognize, assimilate, and exploit externally sourced contributions, for open-source engagement to translate into profitability gains. Second, the role of *VCR* varies across high-profitability configurations, consistent with equifinality. In high-profitability configurations 2 and 4, high external volunteer engagement is present alongside *R&D* and *OSDI*, suggesting a path to high profitability through active community mobilization. In high-profitability configurations 1, 3, and 5, high external engagement is absent, but high *Labor* intensity is present, suggesting an alternative path through a well-resourced internal workforce that compensates for relatively lower external participation. This pathway likely does not run through the *VCR*-amplified labor-productivity mechanism, consistent with the partial mediation results reported in §4.2. Third, low-profitability configurations present a contrasting pattern that exhibits causal asymmetry. *OSDI* is absent in all four low-profitability configurations, indicating that low engagement with open-source development is consistently associated with poor profitability outcomes among adopters. *Capital* and *Industry Concentration* are present across all four low-profitability configurations, while *R&D* is absent in two of the four. This suggests that asset-heavy firms in concentrated industries with limited open-source engagement exhibit relatively low profitability, particularly when they also lack sufficient *R&D* investment. *Capital* is absent from four of the five high-profitability configurations and present in all low-profitability configurations, revealing that *Capital* intensity does not always drive open-source value capture. Finally, *Market Share* is absent from three of the five high-profitability configurations and *Industry Concentration* from all five. Both are present across all low-profitability configurations, suggesting that open-source value capture is associated with less concentrated, more competitive environments rather than with an established market position.

## 5. Discussion

### *5.1 Summary of Results*

This study set out to understand whether open-source product development improves firm profitability, through what mechanism it does so, and under what firm-resource configurations the benefits are realized. Our results provide affirmative and nuanced answers. Firms adopting open-source product development via GitHub realized, on average, a 4%–5% increase in *Gross Margin*, an effect that is robust across multiple specifications. The moderated mediation analysis reveals that this profitability effect operates partially through enhanced *Labor Productivity* and depends on the extent of participation by external contributors. The configurational analysis further demonstrates that open-source engagement translates into profitability through multiple, equifinal resource configurations, with *R&D* intensity functioning as an important factor.

### *5.2 Implications*

Prior research has examined many facets of the open-source phenomenon, yet the pathways from open-source engagement to profitability have remained opaque. Our analysis demonstrates that open-source product development improves the profitability of U.S. high-tech firms. This finding provides practitioners with an evidence-based benchmark for evaluating the financial case for open-source engagement, though the heterogeneity revealed by the configurational analysis cautions against treating it as a universal prescription.

By demonstrating that labor productivity serves as a partial mediating mechanism, we extend the work of Nagle (2019), who showed that *using* open-source software improves productivity for firms with complementary capabilities. Our work provides evidence on the *production* side: firms that actively develop products through open-source communities realize labor productivity gains, and these gains translate into profitability improvements, providing a knowledge-integration-based account grounded in the knowledge-based view of the firm (Kogut and Zander 1992; Grant 1996). More broadly, we contribute to the knowledge-based view by showing that the value a firm derives from leveraging externally distributed knowledge depends on the composition of the contributor base and on the firm's absorptive capacity (cf. Grant and Phene 2022). The moderated mediation model extends the knowledge-based nomological network in the open-source context by identifying *Volunteer Contribution Ratio* as a boundary condition on the labor-productivity pathway. The partial mediation we observe suggests that additional mechanisms beyond labor productivity, such as community governance and responsive engagement (Di Tullio and Staples 2013; Germonprez et al. 2017), licensing-driven value generation (August et al. 2018), or signaling and innovation benefits facilitated by open-source community engagement (Conti et al. 2025), may also contribute to the profitability effect. Disentangling these additional pathways represents a productive avenue for future research.

A practically relevant finding is that the labor-productivity pathway becomes discernibly positive only after external contributors account for a substantial share of development activity, in our sample approximately 35%. Rather than a universal cutoff, this offers managers a practical diagnostic: projects that attract little external participation may be using GitHub primarily as a development tool and are less likely to realize the productivity and profitability benefits that accrue to projects with substantive community engagement. The implication is that investments in community cultivation, spanning responsive governance, process design, and permissive licensing, may be as important as the technical decision to open-source a project (Shaikh and Vaast 2016; August et al. 2018; Lindberg et al. 2024). For researchers, this conditionality finding suggests that the construct of open-source engagement should clearly account for the composition of the contributor base.

The configurational analysis reveals that profitability gains are heterogeneous and depend on complementary resource investments. R&D intensity is present in all high-profitability configurations and absent in half of the low-profitability ones, revealing a clear insight for both theory and practice: open-source engagement is not a substitute for internal R&D investment but a complement to it, consistent with absorptive capacity theory (Cohen and Levinthal 1990). Firms that adopt open-source product development without maintaining robust R&D capabilities risk being unable to absorb, filter, and commercially exploit the external contributions they receive. This pattern is inconsistent with a cost-arbitrage interpretation. If open-source product development created value merely by substituting free external effort for paid internal effort, firms with limited internal knowledge capacity should also capture the benefits; instead, value accrues to firms with the R&D-based absorptive capacity to evaluate, integrate, and commercialize external contributions (Zahra and George 2002; Tang et al. 2020). Thus, technology executives should view open-source engagement and R&D investment as jointly necessary rather than competing budget priorities.

The equifinal configurations further suggest that there are many paths to profitability through open-source engagement. Some firms succeed by combining R&D investment with active community mobilization; others succeed by pairing R&D investment with a strong internal workforce and more selective external engagement. Rather than prescribing a one-size-fits-all strategy, our results suggest that managers should assess their firm's existing resource profile and pursue the engagement model that best leverages their particular strengths. Prior work on governance, controls, and incentives in open-source settings (Di Tullio and Staples 2013; Shah 2006; Lin and Maruping 2022) provides a starting point for tailoring these strategies. For the research community, we expect these findings to shift the debate on value capture from open-source engagement toward a more nuanced position: a generalizable prescription on open-source value capture is unlikely; firms reap benefits only in certain resource configurations; and those beneficial configurations may not be enduring, as they are subject to volatility from competitive dynamics and community-level evolution. This calls for continued inquiry as open-source practices of commercial software firms mature, an agenda that our study advances.

### *5.3 Limitations and Opportunities for Future Research*

We acknowledge several limitations that simultaneously define opportunities for future investigation. First, our sample includes only U.S.-based high-tech firms. Open-source product development is a global phenomenon, and the institutional, regulatory, and labor market conditions that shape its profitability implications may differ across national and industry contexts. Replicating this analysis in other geographies and industrial sectors would help establish the generalizability of our findings. Second, our analysis examines open-source projects directly initiated by the firms in our sample. The product development literature has documented diverse modes of community engagement, from firms contributing to projects they do not host to firms incorporating externally developed components (Chen et al. 2022; Lin and Maruping 2022). Whether the labor productivity mechanism operates similarly across these different engagement modes warrants investigation. Third, our panel structure enables causal identification of the average treatment effect and estimation of the moderated mediation, but it does not fully capture the temporal dynamics of knowledge integration. The benefits of external contributor engagement may take time to materialize as firms develop the governance capabilities and absorptive routines needed to integrate community contributions effectively. Longitudinal designs that model the evolution of the *Volunteer Contribution Ratio* and its lagged effects on productivity would illuminate these dynamics. Fourth, we did not investigate potential interrelationships among the projects hosted on GitHub by the firms in our sample. Firms increasingly pursue value creation through business ecosystems (Agarwal and Kapoor 2023), and it is possible that the interdependencies among a firm's open-source projects, such as shared infrastructure, cross-project knowledge flows, or portfolio-level synergies, influence the profitability effects we observe. Finally, while we focused on *Gross Margin* as a direct and tangible performance measure, there may be other indirect benefits of open-source engagement that our analysis does not capture, including talent acquisition advantages, reputational benefits, and strategic option value. Investigating these broader returns would provide a more complete picture of the value proposition of open-source product development.

### *5.4 Conclusion*

As the software industry's reliance on open-source collaboration deepens, the question of whether this strategy creates business value takes on increasing strategic significance. This study provides evidence that open-source product development improves profitability, but not unconditionally. A key mechanism operates through distributed knowledge integration that enhances labor productivity, is activated by meaningful external contributor participation, and is enabled by the absorptive capacity that R&D investments provide. The implication is that open-source engagement must be designed, resourced, and governed as a knowledge integration strategy suited to a firm's own resource configuration.

**Profitability of Open-Source Software Product Development**

Forthcoming in *Information Systems Research*

**August, 2026**

Shivendu P. Singh
Southern Illinois University
shising@siue.edu

Narayan Ramasubbu
University of Pittsburgh
narayanr@pitt.edu

Chris F. Kemerer
University of Pittsburgh
ckemerer@pitt.edu

M. Zia Hydari
University of Pittsburgh
zia.hydari@pitt.edu

## Online Appendix

## §A: Empirical Studies on Open-Source Software Development

### Table A1: Thematic Review of Studies on Open-Source Software

| **The Firm-Community Interface: Corporate Strategy and Value Creation** | | | |
|---|---|---|---|
| **Study Title (citation)** | **Year** | **Outcome Measure** | **Key Findings** |
| Open at the Core: Moving from Proprietary Technology to Building a Product on Open Source Software (Haese and Peukert 2026) | 2026 | Development activity, security, product quality, market share | A firm's move to an open-source core (Microsoft Edge to Chromium) can increase overall development activity (driven by the firm), improve product quality and release cycles, increase security vulnerability reporting, and lead to gains in market share. |
| Flourish or Perish? The Impact of Technological Acquisitions on Contributions to Open-Source Software (Chen et al. 2022) | 2022 | Internal and external developer contributions | After a firm is acquired, external contributors increase their contributions to the target firm's projects. Internal contributors reduce contributions to the target's projects but increase contributions to other projects in the broader OSS community. |
| Open Source Collaboration in Digital Entrepreneurship (Lin and Maruping 2022) | 2022 | Value of digital startups | Digital startups in the conception and commercialization stages benefit more from inbound OSC (using external OSS), whereas those in the growth stage benefit more from outbound OSC (contributing to external OSS). |
| Opening Up Intellectual Property Strategy: Implications for Open Source Software Entry by Start-up Firms (Wen et al. 2016) | 2016 | New OSS product introductions by start-up firms | A large firm's supportive IP strategy (e.g., a patent non-assertion pledge) stimulates new OSS product entry by startups. This effect is stronger in markets with highly cumulative innovation and concentrated patent ownership. |
| Beyond Free Software: An Exploration of the Business Value of Strategic Open Source (Morgan and Finnegan 2014) | 2014 | Business value creation and capture | Viewing OSS as a value creation process (rather than free software) provides value through access to knowledge and innovation. This requires a strategic shift from ownership to openness and collaboration with external parties. |
| Firms as Incubators of Open-Source Software (Mehra et al. 2011) | 2011 | Optimal employment contracts (principal-agent model) | Firms design incentive contracts (e.g., bonuses) to manage employee effort allocation between proprietary and OSS projects. Typically, firms offer a bonus for only one project type, unless attractive alternative employment opportunities exist, in which case they may offer bonuses for both projects. |
| The Penguin Has Entered the Building: The Commercialization of Open Source Software Products (Fosfuri et al. 2008) | 2008 | Likelihood of a for-profit firm releasing a commercial OSS product | Firms with large stocks of software patents are more likely to release OSS products, while firms with large stocks of software trademarks are less likely to do so. Firms with large stocks of hardware trademarks are more likely to release OSS products. |

**Competition Between Open Source and Proprietary Software**

| Study Title (citation) | Year | Outcome Measure | Key Findings |
|---|---|---|---|
| Competition Among Proprietary and Open-Source Software Firms: The Role of Licensing in Strategic Contribution (August et al. 2021) | 2021 | Quality investments, prices, consumer surplus, social welfare | A more restrictive license can be detrimental to social welfare in markets where firms have similar capabilities. It can be beneficial, encouraging greater effort and quality, when capabilities are dispersed or favor the OSS provider. |
| Competitive Strategy for Open Source Software (Kumar et al. 2011) | 2011 | Product quality, consumer surplus, industry profits | In a market for Commercial Open Source Software (COSS), free-riding behavior is supported in equilibrium. A mandatory sharing setting can lead to high-quality products, and free riding can increase both profits and consumer surplus. |
| A Strategic Analysis of Competition Between Open Source and Proprietary Software (Sen 2007) | 2007 | Market share, profitability of proprietary software vendor | In markets with low network effects, a proprietary software vendor can be more profitable when competing against a commercial OSS offering than as a monopolist. OSS can act as a market-segmenting force. |
| Two-Sided Competition of Proprietary vs. Open Source Technology Platforms and the Implications for the Software Industry (Economides and Katsamakas 2006) | 2006 | Pricing, sales, profitability, social welfare | When a system based on an open source platform competes with a proprietary system, the proprietary system is likely to dominate in both market share and profitability. The variety of applications is greater when the platform is open source. |
| Open Source Software and the "Private-Collective" Innovation Model: Issues for Organization Science (von Hippel and von Krogh 2003) | 2003 | Conceptual model of innovation ("private-collective" model) | Proposes that OSS development is an exemplar of a "private-collective" model of innovation, combining elements of the private investment (proprietary) and collective action models, which can offer society the "best of both worlds" under many conditions. |

**Governance, Control, and Licensing**

| Study Title (citation) | Year | Outcome Measure | Key Findings |
|---|---|---|---|
| Discursive Modulation in Open Source Software: How Online Communities Shape Novelty and Complexity (Lindberg et al. 2024) | 2024 | Software novelty and complexity | Community discourse (filtering and mixing of ideas based on shared doctrines) shapes the novelty and complexity of the software artifact, imprinting community values onto the code. |
| Generating Value Through Open Source: Software Service Market Regulation and Licensing Policy (August et al. 2018) | 2018 | Social welfare, originator's decisions (economic model) | A less restrictive (e.g., BSD-style) license can improve social welfare when a competing contributor is adept at reaping benefits from development. A more restrictive (e.g., GPL-style) license can be optimal for welfare even with a cost-effective contributor if the originator is better at leveraging investment. |
| Collectivism, Creativity, Competition, and Control in Open Source Software Development (Germonprez et al. 2014) | 2014 | Emergent governance structures | OSS communities develop complex, co-existing governance structures beyond the "cathedral vs. bazaar" dichotomy, including meritocracy, adhocracy, and family/republic, reflecting a blend of control, creativity, and collectivism. |
| Determinants of the Choice of Open Source Software License (Sen et al. 2008) | 2008 | Developer license preference | Developer motivations (intrinsic vs. extrinsic) are correlated with preferences for different license types. The choice of license can be used as a strategic tool to attract and cultivate a specific type of developer community. |
| Motivation, Governance, and the Viability of Hybrid Forms in Open Source Software Development (Shah 2006) | 2006 | Developer motivation, project viability | Developer motives evolve from extrinsic need-based participation to intrinsic hobby-like engagement. Governance structures that support this evolution are critical for retaining a core group of "hobbyists" who are essential for the long-term viability and maintenance of the software code. |

**Developer Motivation, Collaboration, and Community Dynamics**

| **Study Title (Citation)** | **Year** | **Outcome Measure** | **Key Findings** |
|---|---|---|---|
| More is not Necessarily Better: An Absorptive Capacity Perspective on Network Effects in Open Source Software Development Communities. (Tang et al. 2020) | 2020 | Knowledge absorption, project success | Network connections provide access to external knowledge, but absorption depends on a project's internal knowledge (absorptive capacity). Knowledge breadth helps absorb knowledge from network depth, and vice versa, but there are inhibiting trade-offs. Bridge members are key to knowledge transfer. |
| Developer Centrality and the Impact of Value Congruence and Incongruence on Commitment and Code Contribution Activity in Open Source Software Communities (Maruping et al. 2019) | 2019 | Developer commitment, code contribution activity | Developer commitment is driven by the reduction of uncertainty. This occurs through passive (congruence between developer and community values) and interactive (centrality in the communication network) avenues. Commitment fully mediates the impact on actual code contributions. |
| Effect of "Following" on Contributions to Open Source Communities (Moqri et al. 2018) | 2018 | Developer contribution level | Gaining social status signals (followers on GitHub) has a significant positive causal effect on a developer's contribution level, especially for freelancers. This links social rewards within the community to future economic opportunities. |
| "Computing" Requirements for Open Source Software: A Distributed Cognitive Approach (Xiao et al. 2018) | 2018 | Requirements for engineering process | OSS requirements engineering is a collective cognitive process where the community "computes" requirements through social interaction and shared artifacts. The process involves three stages: excavation, instantiation, and testing-in-the-wild. |
| Coordinating Interdependencies in Online Communities: A Study of an Open Source Software Project (Lindberg et al. 2016) | 2016 | Management of work interdependencies | OSS communities manage "unresolved interdependencies" (both technical and social) through the emergence of distinct, structured routines, moving beyond simple arm's length coordination mechanisms. |
| Perceived Firm Attributes and Intrinsic Motivation in Sponsored Open Source Software Projects (Spaeth et al. 2015) | 2015 | Intrinsic motivation of volunteer participants | A sponsoring firm's perceived community-based credibility (expertise and trustworthiness) and openness (knowledge exchange) positively impact volunteers' intrinsic motivation, an effect mediated by the volunteer's social identification with the community. |
| All Are Not Equal: An Examination of the Economic Returns to Different Forms of Participation in Open Source Software Communities (Hann et al. 2013) | 2013 | Economic returns (wages) for OSS participants | Participation in OSS communities yields financial rewards. Credentials earned through a merit-based ranking system are associated with up to an 18% wage increase. Project management roles provide additional rewards if the participant's professional job is also in IT management. |
| The Effects of Extrinsic Motivations and Satisfaction in Open Source Software Development (Ke and Zhang 2010) | 2010 | Developer task effort | Differentiates four types of extrinsic motivation (external, introjected, identified, integrated) and shows that more autonomous forms lead to greater effort. This effect is moderated by the satisfaction of needs for competence, autonomy, and relatedness. |

| **Open Source and Team Composition/Dynamics** | | | |
|---|---|---|---|
| **Study Title (citation)** | **Year** | **Outcome Measure** | **Key Findings** |
| Sourcing Knowledge in Open Source Software Projects: The Impacts of Internal and External Social Capital on Project Success (Daniel et al. 2018) | 2018 | OSS project success | Examines how internal social capital (participant diversity in language, role, contribution) and external social capital (project diversity in environment, connectedness) interact. Finds that project connectedness can reduce the positive impact of role and contribution diversity on success. |
| The Impact of Ideology Misfit on Open Source Software Communities and Companies (Daniel et al. 2018) | 2018 | Employee commitment, code contributions | Ideological diversity (misfit) has complex effects. "Under-fit" (employee's ideology is stronger than the group's) decreases commitment to both the company and the community. "Over-fit" (group's ideology is stronger than the employee's) increases company commitment but decreases community commitment. |
| The Impact and Evolution of Group Diversity in Online Open Collaboration (Ren et al. 2016) | 2016 | Group productivity, member withdrawal | Tenure disparity follows a bell curve where moderate variety boosts productivity and retention, while interest variety consistently enhances output throughout a project's life cycle. Groups naturally self-organize toward optimal diversity levels, demonstrating that diverse experiences drive performance and inform better design for collaborative tools. |
| The Effects of Diversity in Global, Distributed Collectives: A Study of Open Source Project Success (Daniel et al. 2013) | 2013 | Project success (level of activity) | Technical diversity has a positive effect on project success, but this effect diminishes at high levels. Geographic diversity has an inverted U-shaped relationship with success; moderate levels are beneficial, but very high levels can be detrimental. |
| Network Effects: The Influence of Structural Capital on Open Source Project Success (Singh et al. 2011) | 2011 | Project success (rate of knowledge creation) | Finds that internal team cohesion positively impacts success. External cohesion and technological diversity of the external network have an inverted U-shaped relationship with success (moderate levels are best). |
| Following the Sun: Temporal Dispersion and Performance in Open Source Software Project Teams. (Colazo and Fang 2010) | 2010 | Development speed, software quality | Temporal dispersion (variation in work hours across time zones) in OSS teams positively affects both the speed of development and the quality of the software produced. |
| Developer Heterogeneity and Formation of Communication Networks in Open Source Software Projects (Singh and Tan 2010) | 2010 | Formation of communication networks | Investigates how diversity among developers (heterogeneity) in dimensions such as skills and experience influences the structure and formation of communication networks within projects. |
| Location, Location, Location: How Network Embeddedness Affects Project Success in Open Source Systems (Grewal et al. 2006) | 2006 | Technical and commercial project success | Network embeddedness (the nature of relationships among projects and developers) has strong but complex effects on project success. Greater embeddedness is not always beneficial; its effect can be positive or negative depending on other factors like project age and visibility. |
| The Impact of Ideology on Effectiveness in Open Source Software Development Teams (Stewart and Gosain 2006) | 2006 | Team trust, communication quality, team effectiveness | A strong, shared ideology among team members leads to a stronger sense of identification with the team. This, in turn, enhances trust and the quality of communication, which are key drivers of overall team effectiveness. |

**The Algorithmic Nature of OSS Work and Coordination**

| **Study Title (citation)** | **Year** | **Outcome Measure** | **Key Findings** |
|---|---|---|---|
| Algorithmic Interactions in Open Source Work (Shaikh and Vaast 2023) | 2023 | Processes of algorithmic interaction, nature of augmentation | Algorithms are essential for OSS work, addressing issues not resolved by modularity or parallel development. They facilitate work through managing, organizing, and supervising processes. Augmentation is a bidirectional relationship: algorithms augment developers' work, and developers augment algorithms' work. |

**Innovation and Open Hardware**

| **Study Title (citation)** | **Year** | **Outcome Measure** | **Key Findings** |
|---|---|---|---|
| From Bits to Atoms: Open Source Hardware at CERN (Priego and Wareham 2023) | 2023 | Development process of a hybrid digital object (hardware and software) | The development of hybrid objects in an open source community is determined by three attributes: embodiment, modularity, and granularity. The assumption that hardware and software require distinct development modes is overly simplistic. |

**Table A2: Review of Empirical Studies on Open-Source Development and Firm Performance**

| Study Title (citation) | Year | Method | Outcome Measure | Key Findings |
|---|---|---|---|---|
| The Transformation of Open Source Software. (Fitzgerald 2006) | 2006 | Conceptual | | The OSS has evolved from a niche phenomenon into a commercially viable "OSS 2.0." This transformation signifies a fundamental alteration of the software industry, moving away from a purely proprietary-driven model. |
| Impact of Internal Open-Source Development on Reuse: Participatory Reuse in Action (Vitharana et al. 2010). | 2010 | Single Case Study | Software Reusability | The use of open-source software leads to the creation of reusable assets in software development. |
| The Impact of Open Source Software on the Strategic Choices of Firms Developing Proprietary Software. (Jaisingh et al. 2008) | 2008 | Analytical Model | Product Quality | A firm may produce lower-quality software when facing an OSS competitor, and the quality of the proprietary software decreases as the quality of the OSS increases. |
| Dynamics of Competition on Openness Strategies and Software Maintenance. (Mallipeddi et al. 2024) | 2024 | Analytical Model | Price and quality of the product | When demand increases with openness, firms should choose an extreme strategy (fully open or fully proprietary), as partial openness is often suboptimal. |
| Open to Your Rival: Competition between Open Source and Proprietary Software under Indirect Network Effects (Wang et al. 2020) | 2020 | Analytical Model | Firm profit | Under different conditions of indirect network effect and software service dependency, open-source software firms may compete with proprietary software firms and generate higher profits. |
| Open Sourcing as a Profit-Maximizing Strategy for Downstream Firms (Gambardella and von Hippel 2019) | 2019 | Analytical Model | Firm Profit | Open sourcing may improve the profits for downstream firms due to scale-stealing from proprietary product firms and reduced cost of manufacturing. |
| Open Collaboration for Innovation: Principles and Performance (Levine and Prietula 2014) | 2014 | Analytical Model | Firm Productivity | Through computational simulations it has been observed that open collaborations are remarkably robust, functioning effectively even in challenging environments characterized by high free-riding, minority cooperation, or rival goods. |
| Open-Source Software and Firm Productivity. (Nagle 2019) | 2019 | Panel Data Analysis | Firm Productivity | A positive and significant impact on value-added productivity only for firms that possess an ecosystem of complementary capabilities (e.g., skilled IT staff, adaptive processes). For firms without these complements, there is no impact. |
| Outsourcing to an Unknown Workforce: Exploring Open Sourcing as a Global Sourcing Strategy (Ågerfalk and Fitzgerald 2008) | 2008 | Case-study and Survey | Human Capital Performance | Open-sourcing enables companies to access a broader talent pool for recruitment and generates qualitative evidence of development efficiency benefits; formal labor productivity effects are not measured. |
| Impact of Open-Source Community on Cryptocurrency Market Price: An Empirical Investigation (Petryk et al. 2023) | 2023 | Longitudinal Analysis | Market Financial Performance | A crypto-currency project's popularity and development on an open-source platform influences its market price. |
| Beyond Free Software: An Exploration of the Business Value of Strategic Open Source (Morgan and Finnegan 2014) | 2014 | Case Study Survey Analysis | Innovation Performance | Most firms observed non-financial gains like knowledge acquisition and idea generation from open-source involvement. |
| External Technology Sourcing and Innovation Performance in LMT Sectors: An Analysis Based on the Taiwanese Technological Innovation Survey (Tsai and Wang 2009) | 2009 | Survey Analysis | Innovation Performance | Inward licensing is not significant predictor of innovation performance; internal R&D weakens returns to R&D outsourcing but contingently changes the returns to different collaboration partners. |
| External Sources of Knowledge, Governance Mode, and R&D Performance (Fey and Birkinshaw 2005) | 2005 | Cross-sectional Analysis | Innovation Performance | External sourcing of knowledge is linked to superior R&D performance, and the link is moderated by openness to new ideas. |
| The Paradox of Openness and Value Protection Strategies: Effect of Extramural R&D on Innovative Performance (Wadhwa et al. 2017) | 2017 | Cross-sectional Analysis | Innovation Performance | External R&D innovation has an inverted U-shaped relationship with the innovation performance of a firm. |
| Scanning the Commons? Evidence on the Benefits to Startups Participating in Open Standards Development (Waguespack and Fleming 2009) | 2009 | Longitudinal Observational Analysis | IPO and Acquisition | The startups that participate in an open standards community have a greater likelihood of an initial public offering or acquisition. |
| Beefing IT Up for Your Investor? Engagement with Open Source Communities, Innovation, and Startup Funding: Evidence from GitHub. (Conti et al. 2025) | 2025 | Longitudinal Observational Analysis | Investor Funding | The early-stage startups that actively engage with OSS communities on platforms like GitHub have a substantially higher likelihood of receiving funding. |
| Software Piracy in the Presence of Open Source Alternatives. (Machado et al. 2017) | 2017 | Analytical Model | Software Price and Piracy Control | The appearance of an OSS alternative leads the incumbent to reduce both its price and its level of piracy control. Piracy can also limit the incumbent's losses by maintaining network effects against the OSS competitor. |

## §B: Measuring a Firm's Open-Source Development

We identified high-tech firms' adoption and use of open-source software product development through the following steps.

### 1. Identifying firms in the high-tech sector:

We gathered a list of firms in the U.S. high-tech industry sector using the following 3-digit and 6-digit codes of the North American Industry Classification System (NAICS). These codes were selected based on prior studies of the high-tech industry published by the Bureau of Labor Statistics (BLS) and the U.S. Census Bureau (Wolf and Terrell 2016; Goldschlag and Miranda 2020), as well as the Information Technology and Innovation Foundation (Atkinson 2022). The industry sectors listed below each contained at least one firm involved in open-source software product development during 2001–2025.

- 334: Computer and electronic product manufacturing
- 423430: Computer and Computer Peripheral Equipment and Software Merchant Wholesalers
- 511210: Software Publishers
- 517: Telecommunications
- 518210: Data Processing, Hosting, and Related Services
- 519130: Internet Publishing and Broadcasting and Web Search Portals
- 541: Professional, scientific, and technical services

### 2. Identifying firms' presence on GitHub

For every high-tech firm collected in the previous step, we conducted a thorough Internet search to discover whether the firm had a presence on GitHub or any other open-source software development platform, such as GitLab, SourceForge, and Bitbucket. All open-source adopting firms in our search results maintained a presence on GitHub. Utilizing the search results, we identified the firm's official homepage on GitHub, and we verified the authenticity of the firm's GitHub profile by examining the profile content, official contact email addresses, and the descriptions of other policy content posted by the firm. For example, an initial search using the keywords "Netflix open source" yields the results shown in Figure B1. From these results, we captured Netflix's official GitHub homepage at: https://github.com/netflix, a snapshot of which is shown in Figure B2.

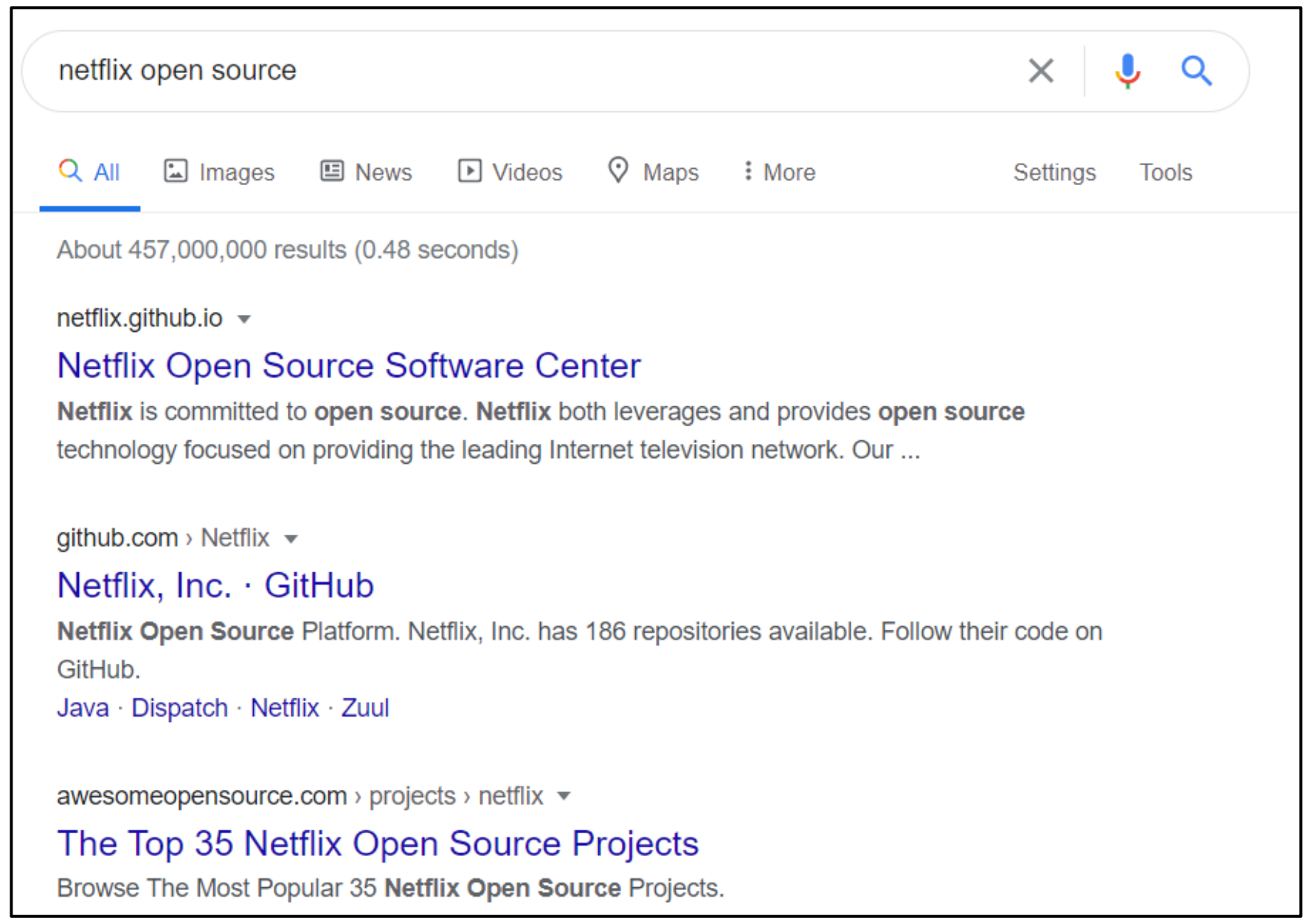

**Figure B1. Search Results using "Netflix open source"**

In addition to the search-based identification of firms' presence on GitHub, we also exploited a commonly utilized URL structure for an organization's GitHub home page, namely,
https://github.com/*NameOfTheFirm*. For each firm in our list, we "pinged" these URLs, and for the ones returning a successful ping test, we examined the contents of their homepages. Additional validity checks were still required, as some firms' official pages either followed a URL structure different from the common pattern or closely resembled one another. For example, the official GitHub page of AVG Technologies is https://github.com/avgtechnologies and not https://github.com/avg, even though both pages are active on GitHub.

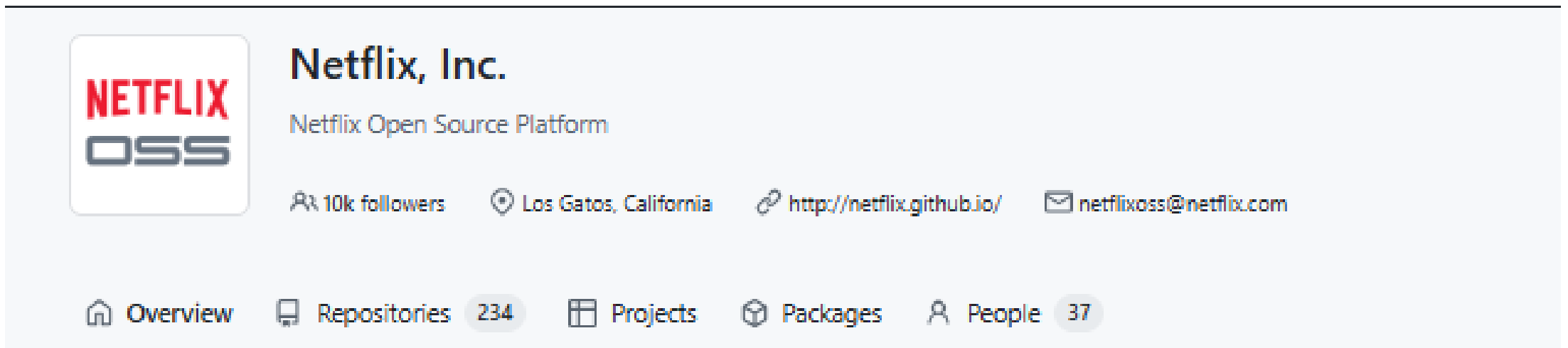

**Figure B2. The official GitHub homepage of Netflix**

## 3. Extracting Firm-level data using GitHub API

Once we identified the URLs of the official GitHub homepages of high-tech firms in our list, we utilized the GitHub API and the command-line utility, CURL, for extracting project-level information. To execute this, the original homepage URLs of firms had to be modified for use with the GitHub API. For example, while the official GitHub homepage of Netflix is https://github.com/netflix, the URL for data extraction through the API is https://api.github.com/orgs/netflix/repos. Figure B3 shows a snapshot of the API output for a query about Netflix's projects.

https://api.github.com/orgs/netflix

Pretty-print

```
{
  "login": "Netflix",
  "id": 913567,
  "node_id": "MDEyOk9yZ2FuaXphdGlvbjkxMzU2Nw==",
  "url": "https://api.github.com/orgs/Netflix",
  "repos_url": "https://api.github.com/orgs/Netflix/repos",
  "events_url": "https://api.github.com/orgs/Netflix/events",
  "hooks_url": "https://api.github.com/orgs/Netflix/hooks",
  "issues_url": "https://api.github.com/orgs/Netflix/issues",
  "members_url": "https://api.github.com/orgs/Netflix/members{/member}",
  "public_members_url": "https://api.github.com/orgs/Netflix/public_members{/member}",
  "avatar_url": "https://avatars.githubusercontent.com/u/913567?v=4",
  "description": "Netflix Open Source Platform",
  "name": "Netflix, Inc.",
  "company": null,
  "blog": "http://netflix.github.io/",
  "location": "Los Gatos, California",
  "email": "netflixoss@netflix.com",
  "twitter_username": null,
  "is_verified": false,
  "has_organization_projects": true,
  "has_repository_projects": true,
  "public_repos": 234,
  "public_gists": 0,
  "followers": 10045,
  "following": 0,
  "html_url": "https://github.com/Netflix",
  "created_at": "2011-07-13T20:20:01Z",
  "updated_at": "2026-04-16T17:57:14Z",
  "archived_at": null,
  "type": "Organization"
}
```

**Figure B3. GitHub API Output for a Query on Netflix's Projects**

Finally, using Google BigQuery, we collected project-level data, including license information, fan-in and fan-out fork counts, and the productive events (e.g., pull requests, commits, pull-request reviews, and issues). We note a few challenges and restrictions that we encountered during the extraction of project-level information through the GitHub API. For example, even with an authentication key, the GitHub API allows only 5,000 extractions per hour. More restrictively, when a firm has more than 100 projects on GitHub, we had to create page-wise URLs for data extraction for each firm. Moreover, for a few firms with large repositories on GitHub, in certain calendar years, we were unable to download complete historical information on their projects, even with an authentication key and modified, page-wise URLs. Restrictive changes in the project permissions over time during our observation period (2001–2025) may be the underlying reason for such errors.

## §C: Comparison of High-Tech Firms in the Sample

Figure C1 shows the sample distribution in panel view. Each row uniquely identifies a firm in the graph. The firms shaded in orange never adopted GitHub. The firms that adopted GitHub are shown in green and blue. The green color indicates a period before adoption, and the blue color after the adoption. Missing year observations are shown in white color. All the adopter firms are grouped in the upper half of the graph, and non-adopters in the bottom half. As is visible, firms adopted GitHub across various years in the timelines, and the panel is unbalanced across both adopter and non-adopter firms, revealing left-censoring, right-censoring, and missing-at-random patterns.

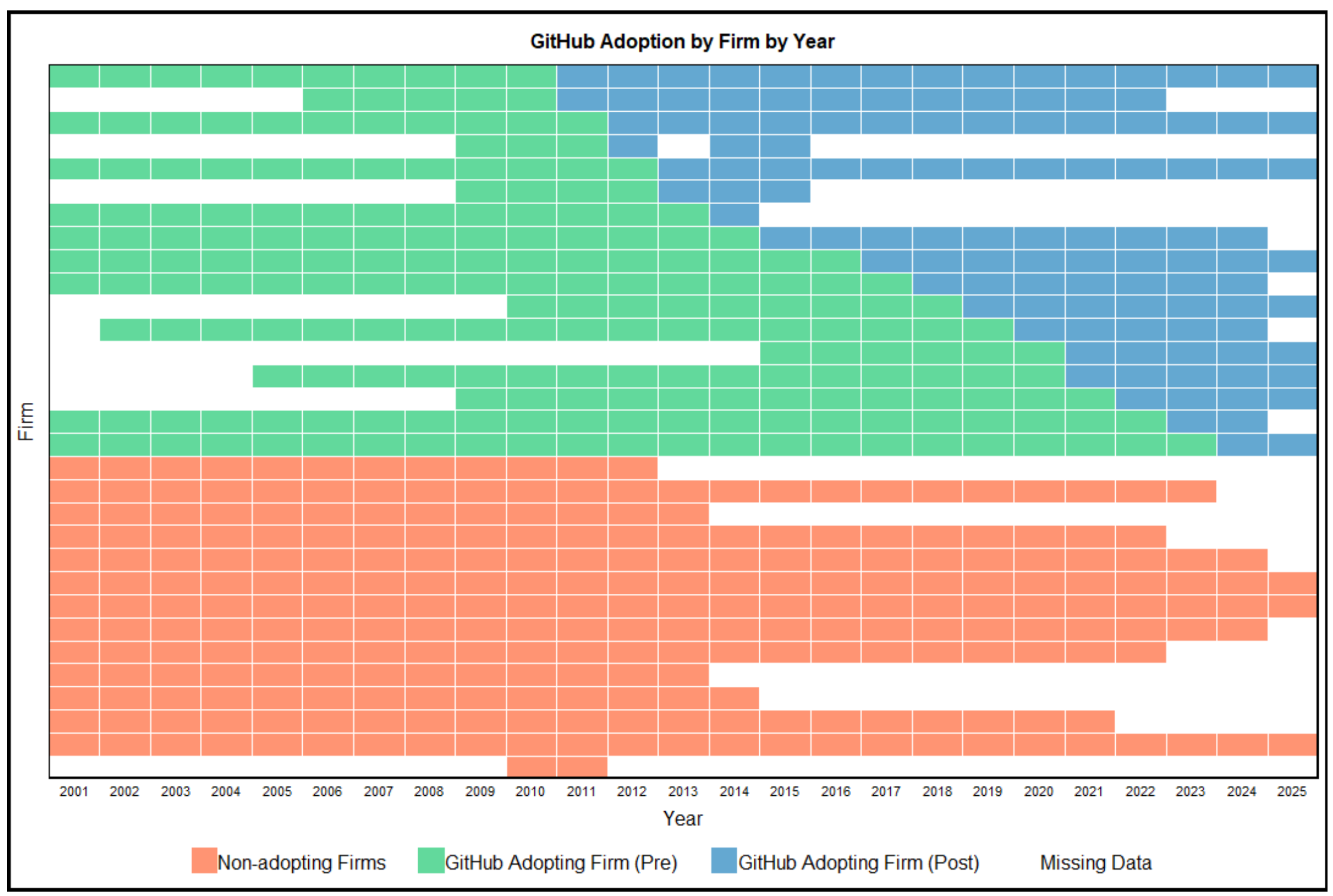

**Figure C1: PanelView of Firms**

Figure C2 compares the distribution of the outcome variable, *Gross Margin*, between the control and treated group firms during the pre- and post-adoption time periods. As seen in Figure C2, the distribution of *Gross Margin* for firms that adopted GitHub is skewed right, relative to their own pre-adoption period as well as to the firms that never adopted GitHub, which suggests a potential positive effect of open-source product development process on the profitability of the treated firms.

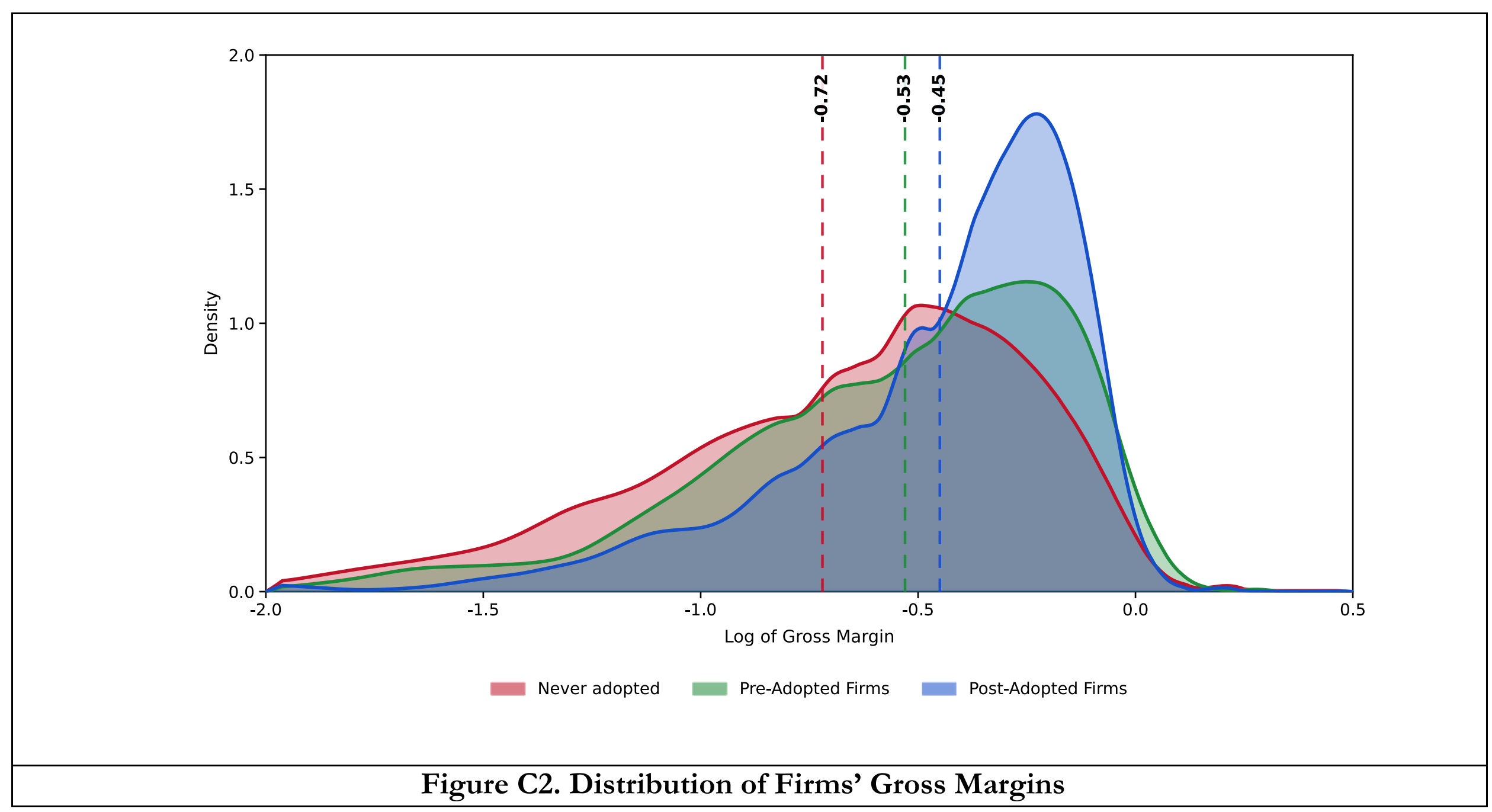

**Figure C2. Distribution of Firms' Gross Margins**

**Table C1. Variables in Dataset**

| Variable | Definition | Source |
|---|---|---|
| *Open-Source Development* | A binary variable denoting a firm that was engaged in open-source product development. | GitHub |
| *Open-Source Development Intensity (OSDI)* | Log of the total number of open-source projects that are actively developed by a firm in a year on GitHub. | GitHub |
| *Volunteer Contribution Ratio (VCR)* | The number of unique observed external (non-employee) contributions divided by total observed contributions (external + internal) across a firm's repositories in a year. | GitHub |
| *Total Forks* | Total number of forks created from all the repositories hosted by a firm in a year on GitHub. | GitHub |
| *Total Stars* | Total number of stars given to all the repositories hosted by a firm in a year on GitHub. | GitHub |
| *Total Issues* | Total number of issues in all the repositories hosted by a firm in a year on GitHub. | GitHub |
| *Gross Margin* | Log of ((*Total Sales* - *Cost of Goods Sold*)/ *Total Sales*) for a firm in a year. | Compustat |
| *Labor Productivity* | Log of Sales/Employee for a firm in a year. | Compustat |
| *R&D*† | Log of R&D investment in millions of dollars (calculated using the perpetual inventory method: 15% depreciation rate) for a firm in the year after deflation. | Compustat |
| *Capital*† | Log of net capital (calculated using the perpetual inventory method: 5% depreciation rate) in millions USD for a firm in a year after deflation. | Compustat |
| *Employees/Labor*† | Log of the number of employees in thousands for a firm in a year. | Compustat |
| *Large Firm* | A binary variable coded as 1 if the firm is in the top 25th percentile of firms in its industry sector based on *Total Sales* in a year, 0 otherwise. | Compustat |
| *Market Share* | The percentage of total sales of a firm by the total sales of the entire sector defined at the NAICS-2 level in a given year. | Compustat |
| *Industry Concentration* | Herfindahl-Hirschman Index of the industry sector of a firm in a year. | Compustat |

*Note:* † In the configurational (fsQCA) analysis (§3.3 and §4.3), *R&D*, *Capital*, and *Labor* are expressed as revenue-normalized intensities and then calibrated into fuzzy-set membership scores ranging from 0 to 1.

Table C2 illustrates a strategic shift in the technological landscape between 2011 and 2025, specifically regarding the adoption of GitHub by industry leaders. In 2011, the control group comprised dominant hardware and software firms such as Apple Inc., Microsoft Corp., and Dell Technologies Inc., which had not yet integrated GitHub into their development workflows. However, by 2025, these firms transitioned into the treated group, becoming participants in the GitHub ecosystem. This mainstream adoption by former control-group entities is reflected in the revenue growth of the top 10 treated firms, whose average sales rose from $5.5 billion USD to $103.47 billion USD as the industry's largest revenue-generating players embraced open-source software product development.

**Table C2: Comparison of Treated and Control Groups†**

| | **2011** | **2025** |
|---|---|---|
| Average *Sales:* Top 10 treated firms | 5.50 billion USD | 103.47 billion USD |
| Average *Sales:* Top 10 control firms | 64 billion USD | 72.83 billion USD |
| Average *Sales* of all treated firms | 3.80 billion USD | 12.04 billion USD |
| Average *Sales* of all control firms | 2.20 billion USD | 7.41 billion USD |
| Top-10 firms in the treated group | Alphabet Inc., Blackberry Ltd., Bsquare Corp., Facebook Inc., Leaf Group Ltd., LinkedIn Corp., Medidata Solutions Inc., Mercadolibre Inc., NVIDIA Corp., and Rentrak Corp. | Apple Inc., Alphabet Inc., NVIDIA, Microsoft Corp., IBM, Dell Technologies Inc., Intel Corp., Cisco Systems Inc., Meta Inc., Hewlett Packard Enterprise. |
| Top-10 firms in the control group | Apple Inc., Cisco Systems, Dell Technologies Inc., Fujitsu Ltd., Intel Corp., Microsoft Corp., NEC Corp., Nippon Telegraph & Telephone, Nokia Corp., Orange. | Techdata Corp., Synnex Corp., Echostar Corp., Block Inc., Amkor Technology Inc., Cadence Design Systems Inc., Paycom Software Inc., I3 Technologies Inc., Fidelity National Info Services, Qwest Corp. |

*Note:* † All USD values are deflated to 2012 values. For multinational firms, we utilize data reported by their U.S. subsidiaries.

## §D: Additional Models and Robustness Results

To ensure the stability of our Stage-1 causal effect estimates and mitigate the influence of extreme observations, we tested the model using samples Winsorized at both the 1% and 5% threshold levels. The results across these specifications, presented in Table D1, remain consistent with those reported in the main text. This shows that the performance gains associated with open-source engagement are consistent and robust to different Winsorization thresholds. Figure D1 presents a cumulative event-study plot using the staggered Difference-in-Differences (DID) estimates of the effect of *Open-Source Development* on *Gross Margin*.

Next, we used the DID estimator proposed by de Chaisemartin and D'Haultfœuille (2026), which accommodates intertemporal treatment reversals. The average effect of GitHub adoption on *Gross Margin* is positive and significant (β=0.031, S.E.=0.014). The test of the joint null of the placebos is insignificant (p=0.358), supporting the validity of the parallel trends and no-anticipation assumptions. These estimation results are presented in Table D2, and Figure D2 visualizes these results for the entire sample. The results are consistent with those reported in the main text.

| **Table D1. Open-Source Product Development and Profitability**<br>*Staggered DID (Callaway and Sant'Anna, 2021)* | | |
|---|---|---|
| Dependent Variable = *Gross Margin* | | |
| Independent Variable | Staggered DID full sample<br>*(Winsorized at 1%)*<br>*Model 1* | Staggered DID full sample<br>*(Winsorized at 5%)*<br>*Model 2* |
| Open-Source Development (ATT) | .047*** (.014) | .048*** (.013) |
| ATT by Group | .043*** (.014) | .045*** (.013) |
| ATT by Calendar Year | .044*** (.015) | .045*** (.014) |
| Control Variables† | Yes | Yes |
| Firm Fixed Effect | Yes | Yes |
| Industry Fixed Effect | Yes | Yes |
| Time Fixed Effect | Yes | Yes |
| Observations | 11,797 | 11,797 |
| Total No. of firms | 977 | 977 |
| No. of treated firms | 231 | 231 |
| Parallel-trend Assumption | 32.111 *(p=1.000)* | 78.177 *(p=1.000)* |

*Note:* Standard errors are in parentheses; *** $p<.01$, ** $p<.05$, * $p<.1$; Control variables included *R&D*, *Capital*, *Employees*, *Large Firm*, *Market Share,* and *Industry Concentration*. The csdid estimator (Callaway and Sant'Anna 2021) utilizes control variables to adjust for conditional parallel trends via inverse probability weighting and/or outcome regression. Consequently, it does not produce or report standalone linear coefficients for these covariates.

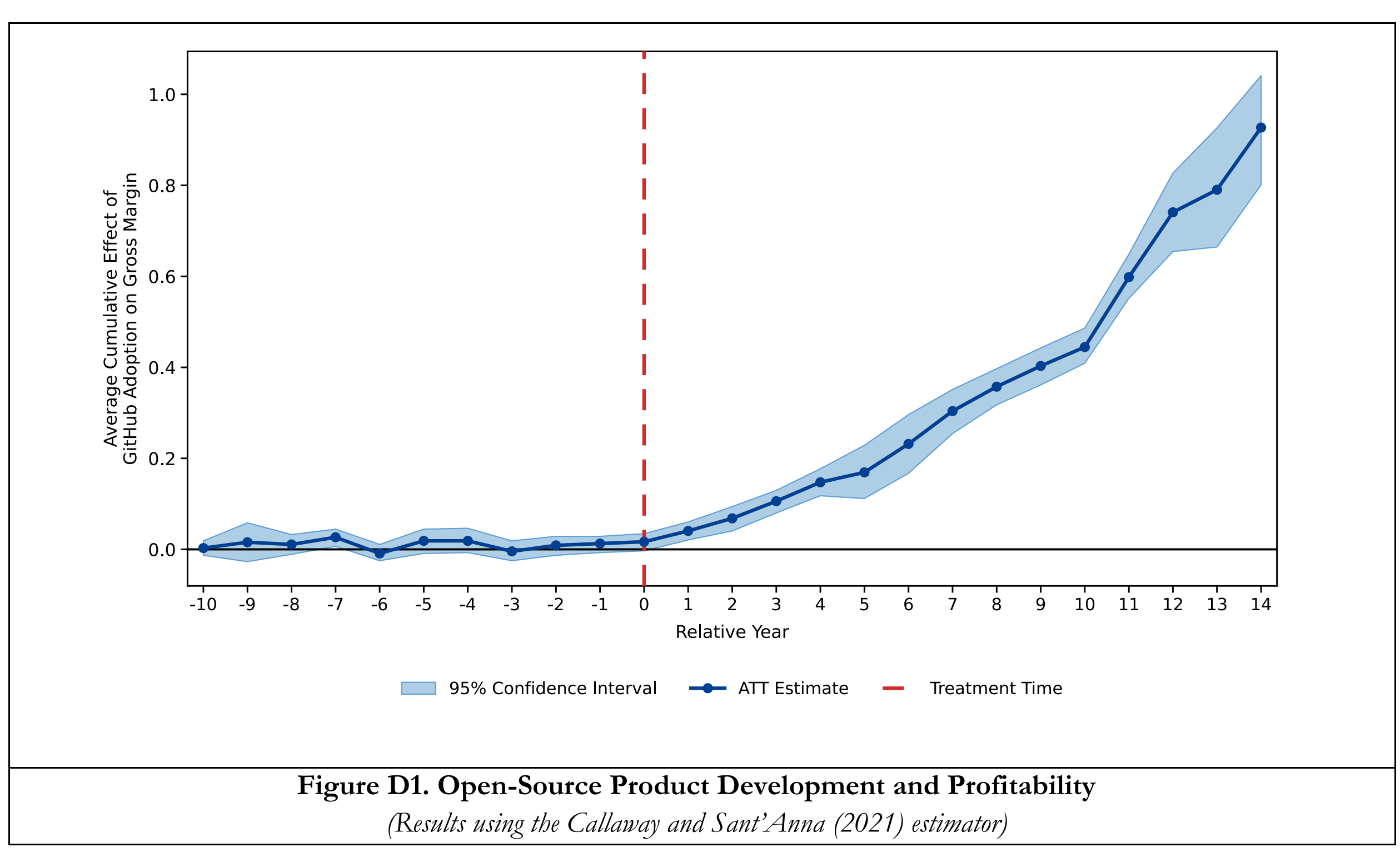

**Figure D1. Open-Source Product Development and Profitability**

*(Results using the Callaway and Sant'Anna (2021) estimator)*

| **Table D2. Open-Source Product Development and Profitability**<br>*Staggered DID (de Chaisemartin and D'Haultfœuille 2026)* | | | |
|---|---|---|---|
| Independent Variable | *Gross Margin* | *Gross Margin (Winsorized at 1%)* | *Gross Margin (Winsorized at 5%)* |
| Open-Source Development (ATT) | .031** (.014) | .033** (.013) | 0.036*** (.012) |
| Control Variables† | Yes | Yes | Yes |
| Firm Fixed Effect | Yes | Yes | Yes |
| Industry Fixed Effect | Yes | Yes | Yes |
| Time Fixed Effect | Yes | Yes | Yes |
| Observations | 5,769 | 5,769 | 5,769 |
| Parallel-trend Assumption | .358 | .269 | .382 |

*Note:* Standard errors are in parentheses; *** $p<.01$, ** $p<.05$, * $p<.1$; Control variables included *R&D*, *Capital*, *Employees*, *Large Firm*, *Market Share,* and *Industry Concentration*. The sample size for this analysis is reduced to 5,769 observations because the staggered DID package for the de Chaisemartin and D'Haultfœuille (2026) estimator allows estimation for only seven time periods after the adoption of the open-source software development process. The p>0.05 on the parallel trend assumption test indicates that there is insufficient evidence to reject the null hypothesis of parallel pre-trends.

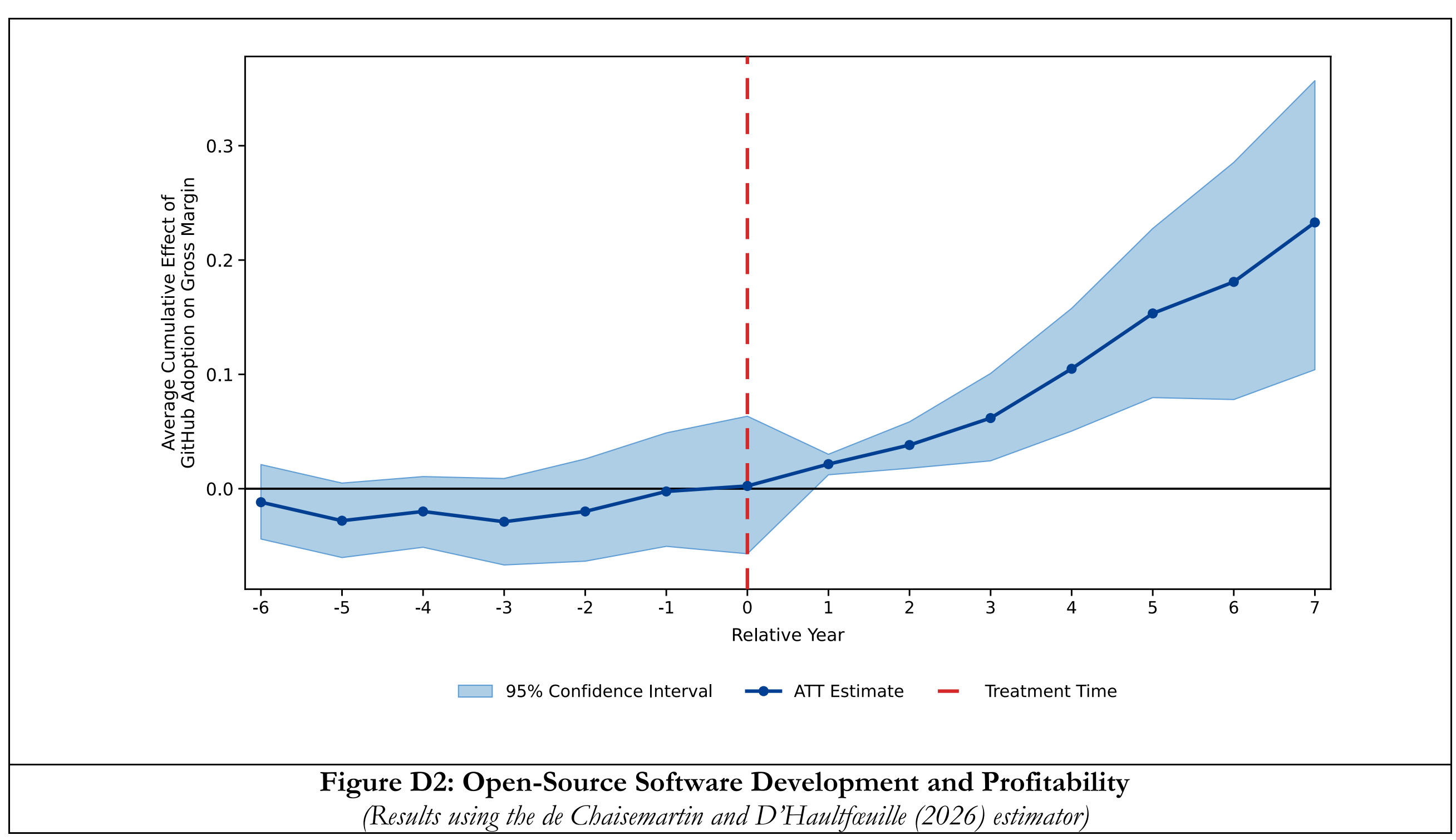

**Figure D2: Open-Source Software Development and Profitability**
*(Results using the de Chaisemartin and D'Haultfœuille (2026) estimator)*

Next, we estimated the effect using the generalized synthetic control method (Xu 2017), which provides a flexible approach for constructing counterfactuals and relaxes the strict parallel trends assumption, allowing for heterogeneous treatment effects across multiple treated units. While we did not observe violations of parallel trends in the DID estimators, an alternative approach that does not rely on this assumption helps build confidence in results (Baker et al. 2022). The generalized synthetic control estimate is consistent with the DID results (ATT=0.053, S.E.=0.020). These results, together with Winsorized specifications, are reported in Table D3 and visualized in Figure D3.

| **Table D3. Open-Source Product Development and Profitability**<br>*Generalized Synthetic Control Method (Xu 2017)* | | | |
|---|---|---|---|
| Independent Variable | *Gross Margin* | *Gross Margin (Winsorized at 1%)* | *Gross Margin (Winsorized at 5%)* |
| Open-Source Development (ATT) | .053** (.020) | .054*** (.020) | 0.051*** (.019) |
| Control Variables† | Yes | Yes | Yes |
| Firm Fixed Effect | Yes | Yes | Yes |
| Industry Fixed Effect | Yes | Yes | Yes |
| Time Fixed Effect | Yes | Yes | Yes |
| *Note:* Standard errors are in parentheses*; *** p<.01, ** p<.05, * p<.1;* Control variables included *R&D, Capital*, *Employees*, *Large Firm*, *Market Share* and *Industry Concentration*. | | | |

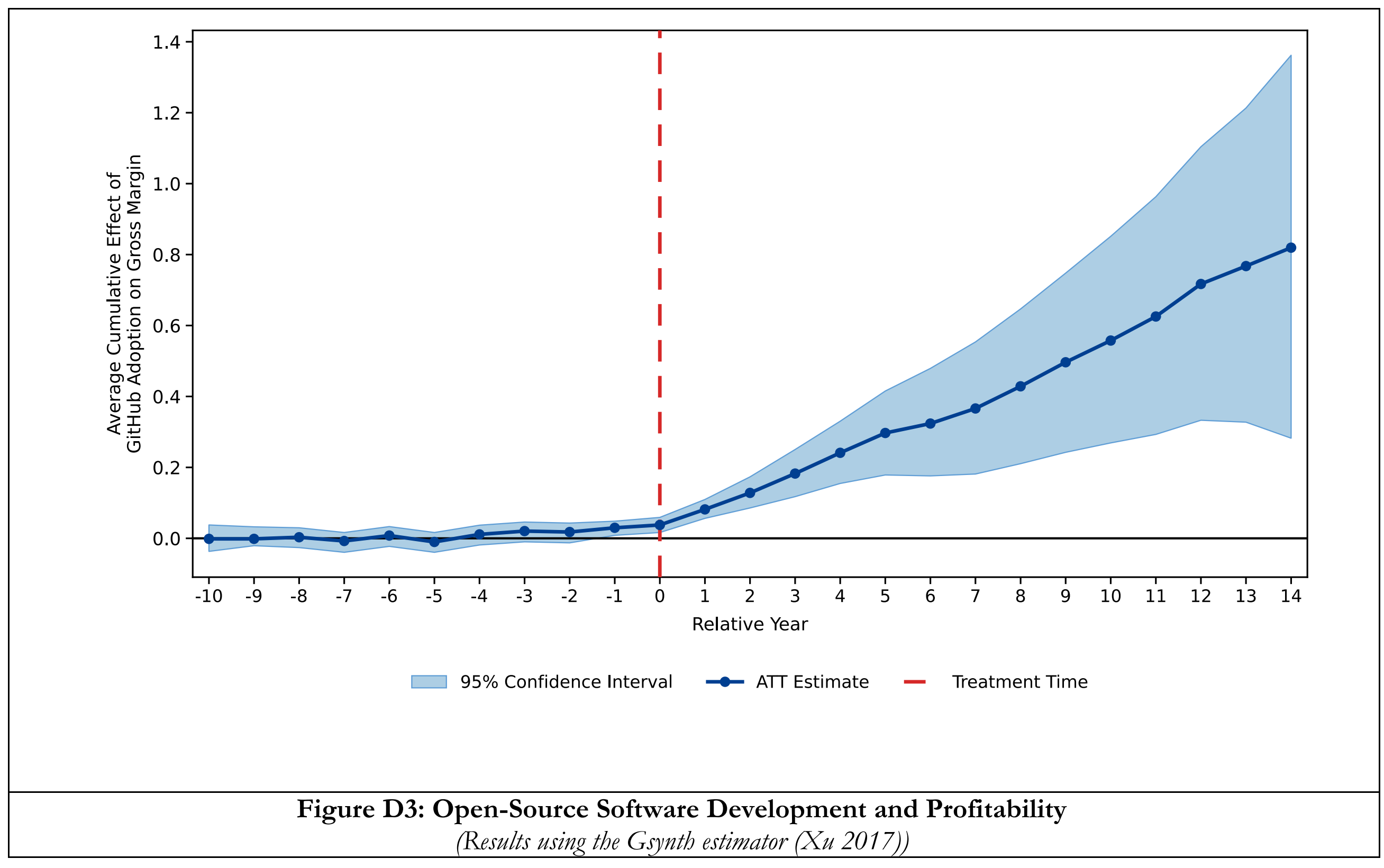

**Figure D3: Open-Source Software Development and Profitability**
*(Results using the Gsynth estimator (Xu 2017))*

Finally, we conducted an instrumental variable analysis using two-stage least squares (2SLS) to address endogeneity concerns due to potential unobserved, time-varying confounders. Following prior studies that use peer adoption to instrument for focal firm technology adoption (Bendig et al. 2023; Gao et al. 2026), we utilized industry-level peer adoption of GitHub as the instrument. We calculated the peer adoption rate at both the 2-digit NAICS and SIC industry levels. In the first stage, the coefficients on peer adoption rate are significant at both the NAICS-2 level ($\beta$=0.282, $p<0.01$) and the SIC level ($\beta$=0.297, $p<0.01$), confirming the relevance of the instrument. In the second stage, the effect of open-source product development on *Gross Margin* remains positive and significant ($\beta$=0.085, $p<0.01$ using NAICS-2; $\beta$=0.084, $p<0.01$ using SIC). These results, which rely on the exclusion restriction, provide further support for the causal interpretation of our DID estimates. Full results are presented in Table D4.

| **Table D4. Open-Source Product Development and Profitability**<br>*Instrumental Variable Analysis* | | | | |
|---|---|---|---|---|
| Independent Variable | Instrument = *Peer adoption rate at NAICS-2 level* | | Instrument = *Peer adoption rate at SIC level* | |
| | Y = *Open-Source Development* | Y = *Gross Margin* | Y = *Open-Source Development* | Y = *Gross Margin* |
| Peer adoption rate (NAICS2) | .282*** (.057) | | | |
| Peer adoption rate (SIC) | | | .297*** (.031) | |
| Open-Source Development | | .085*** (.032) | | 0.084*** (.031) |
| Control Variables† | Yes | Yes | Yes | Yes |
| Firm Fixed Effect | Yes | Yes | Yes | Yes |
| Industry Fixed Effect | Yes | Yes | Yes | Yes |
| Time Fixed Effect | Yes | Yes | Yes | Yes |
| Observations | 11,140 | | 10,951 | |
| Total No. firms | 956 | | 939 | |
| *Note:* Standard errors are in parentheses*; *** p<.01, ** p<.05, * p<.1;* Control variables for first stage included *R&D*, *Capital*, *Employees*, *Large Firm*, *Market Share* and *Industry Concentration*. | | | | |

Table D5 presents the second-stage results of our instrumental variable analysis with *Open-Source Development* as our independent variable and *Gross Margin* as the dependent variable, using Winsorized values to assess robustness to outliers, with sensitivity tests at the 1% and 5% thresholds.

| **Table D5. Open-Source Product Development and Profitability**<br>*Instrumental Variable Analysis (Winsorized)* | | | | |
|---|---|---|---|---|
| Independent Variable | Instrument = *Peer adoption rate at NAICS-2 level* | | Instrument = *Peer adoption rate at SIC level* | |
| | Y = *Gross Margin (Winsorized at 1%)* | Y = *Gross Margin (Winsorized at 5%)* | Y = *Gross Margin (Winsorized at 1%)* | Y = *Gross Margin (Winsorized at 5%)* |
| Open-Source Development | .079*** (.028) | .102*** (.024) | .076***(.028) | 0.096*** (.023) |
| Control Variables† | Yes | Yes | Yes | Yes |
| Firm Fixed Effect | Yes | Yes | Yes | Yes |
| Industry Fixed Effect | Yes | Yes | Yes | Yes |
| Time Fixed Effect | Yes | Yes | Yes | Yes |
| Observations | 11,140 | 11,140 | 10,951 | 10,951 |
| Total No. firms | 956 | 956 | 939 | 939 |
| *Note:* Standard errors are in parentheses*; *** p<.01, ** p<.05, * p<.1;* Control variables for the first stage included *R&D*, *Capital*, *Employees*, *Large Firm*, *Market Share,* and *Industry Concentration*. | | | | |

## §E: Open-Source Development Intensity

To assess whether Microsoft's 2018 acquisition of GitHub affected the relationship between *Open-Source Development Intensity (OSDI)* and *Gross Margin*, we analyzed the pre-acquisition subsample. Table E1 presents results from three specifications: two-way fixed effects (TWFE, Model 1), Arellano-Bond (ABOND, Model 2), and Blundell-Bond (BBOND, Model 3). The results reveal a consistent positive effect of *OSDI* on *Gross Margin* of the firms. To further test the influence of extreme outliers on the *OSDI* and *Gross Margin* relationship, we estimated the ABOND and BBOND models for the pre-acquisition subsample with Winsorized values at 1% and 5% thresholds. The results are presented in Table E2. The effect of *OSDI* on *Gross Margin* is not sensitive to restricting outliers in the dataset. Finally, we repeated the analysis using the full sample, Winsorized at both the 1% and 5% threshold levels. Across all specifications and at both Winsorization levels, the positive effect of *OSDI* on *Gross Margin* remains statistically significant, providing evidence that our findings are robust.

**Table E1. Open-Source Development Intensity and Profitability—pre-Microsoft acquisition of GitHub**

Dependent Variable = *Gross Margin*

| | (1) | (2) | (3) |
|---|---|---|---|
| Independent Variable | TWFE | ABOND | BBOND |
| Open-Source Development Intensity | .019***(.007) | .010***(.003) | .011*** (.003) |
| Control Variables† | Yes | Yes | Yes |
| Firm Fixed Effect | Yes | Yes | Yes |
| Industry Fixed Effect | Yes | Yes | Yes |
| Time Fixed Effect | Yes | Yes | Yes |
| Observations | 8,935 | 7,233 | 8,252 |
| Total No. of firms | 973 | 849 | 940 |

*Note:* Standard errors are in parentheses*; *** p<.01, ** p<.05, * p<.1;* the number of observations in models (2) and (3) differ from those in model (1) due to the use of lagged instrumental variables and unavailability of data for certain lagged calendar years*;* Control variables included *R&D*, *Capital, Employees*, *Large Firm*, *Market Share* and *Industry Concentration*.

**Table E2. Open-Source Development Intensity and Profitability—pre-Microsoft acquisition of GitHub (Winsorized)**

Dependent Variable = *Gross Margin*

| | *(Winsorized at 1%)* | | *(Winsorized at 5%)* | |
|---|---|---|---|---|
| | (1) | (2) | (3) | (4) |
| Independent Variable | ABOND | BBOND | ABOND | BBOND |
| Open-Source Development Intensity | .010*** (.003) | .011*** (.003) | .008*** (.002) | .009***(.002) |
| Control Variables† | Yes | Yes | Yes | Yes |
| Firm Fixed Effect | Yes | Yes | Yes | Yes |
| Industry Fixed Effect | Yes | Yes | Yes | Yes |
| Time Fixed Effect | Yes | Yes | Yes | Yes |
| Observations | 7,233 | 8,252 | 7,233 | 8,252 |
| Total No. of firms | 849 | 940 | 849 | 940 |

*Note:* Standard errors are in parentheses*; *** p<.01, ** p<.05, * p<.1;* Control variables included *R&D*, *Capital, Employees*, *Large Firm*, *Market Share* and *Industry Concentration*.

**Table E3. Open-Source Development Intensity and Profitability—Full sample** *(Winsorized)*

Dependent Variable = *Gross Margin*

| | *(Winsorized at 1%)* | | *(Winsorized at 5%)* | |
|---|---|---|---|---|
| | (1) | (2) | (3) | (4) |
| Independent Variable | ABOND | BBOND | ABOND | BBOND |
| Open-Source Development Intensity | .009*** (.002) | .015*** (.002) | .007*** (.002) | .010***(.002) |
| Control Variables† | Yes | Yes | Yes | Yes |
| Firm Fixed Effect | Yes | Yes | Yes | Yes |
| Industry Fixed Effect | Yes | Yes | Yes | Yes |
| Time Fixed Effect | Yes | Yes | Yes | Yes |
| Observations | 10,095 | 11,154 | 10,095 | 11,154 |
| Total No. of firms | 908 | 957 | 908 | 957 |

*Note:* Standard errors are in parentheses*; *** p<.01, ** p<.05, * p<.1;* Control variables included *R&D*, *Capital, Employees*, *Large Firm*, *Market Share* and *Industry Concentration*.

## §F: Moderated Mediation Mechanism

Figure F1 shows the overall *Volunteer Contribution Ratio* (*VCR*) density in our sample. The *VCR* averages 0.51, with a median of 0.54, indicating that about half of total contributions typically come from volunteers. Most observations fall between 0.41 (25th percentile) and 0.64 (75th percentile), showing moderate dispersion around the center.

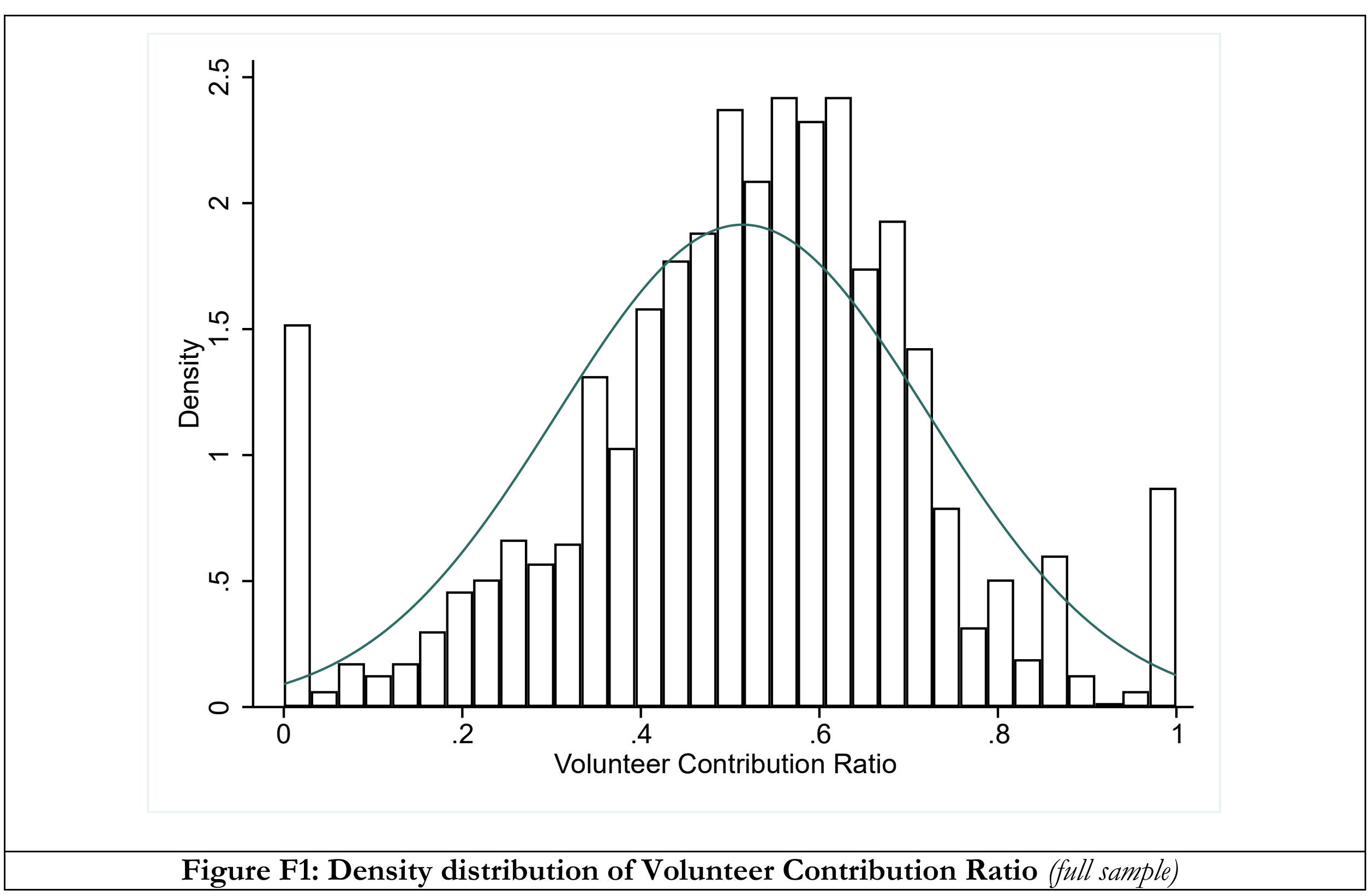

**Figure F1: Density distribution of Volunteer Contribution Ratio** *(full sample)*

To further validate the robustness of our moderated mediation model, we conducted several subsample and sensitivity analyses. First, we re-estimated the moderated mediation model (IV – *OSDI*, moderator – *VCR*, mediator – *Labor Productivity, DV – Gross Margin*) using a pre-acquisition subsample Winsorizing at the 1% and 5% thresholds to mitigate the influence of extreme outliers. The results of these tests are presented in Table F1 – Full pre-acquisition subsample (Model 1), Winsorized at 1% (Model 2), and Winsorized at 5% (Model 3). Across these various specifications, the results remain consistent. *VCR* moderates the effect of *OSDI* on *Gross Margin* mediated by *Labor Productivity*.

The conditional indirect mediated effect of OSDI on *Gross Margin* is visualized in Figure F2 for the pre-acquisition subsample. In this subsample, the mediated effect becomes discernibly positive at a higher level of external contribution (>50%), relative to the full sample. This illustrates the sample-specific nature of the threshold. Finally, we ran the moderated mediation analysis on the full sample with Winsorized variables at 1% and 5% threshold levels. Those results are presented in Table F2, and they are consistent with the non-Winsorized results.

| **Table F1: Open-Source Development Intensity, Labor Productivity, and Profitability**<br>*Moderated Mediation Analysis using VCR (pre-Microsoft acquisition of GitHub sample)* | | | | |
|---|---|---|---|---|
| **DV** | **Predictor** | **Model 1** | **Model 2** (*Winsorized at 1%*) | **Model 3** (*Winsorized at 5%*) |
| Labor Productivity | Constant | 2.412*** (0.277) | 2.711*** (0.258) | 3.176*** (0.232) |
| | Open-Source Development Intensity (*OSDI*) | -0.006 (0.012) | -0.004 (0.011) | -0.001 (0.010) |
| | Volunteer Contribution Ratio (*VCR*) | -0.056 (0.099) | -0.053 (0.092) | -0.040 (0.083) |
| | *OSDI* X *VCR* | 0.121** (0.053) | 0.124** (0.049) | 0.135*** (0.044) |
| | Control Variables† | Yes | Yes | Yes |
| | Year Fixed Effect | Yes | Yes | Yes |
| Gross Margin | Constant | 0.031 (0.145) | -0.624*** (0.121) | -0.785*** (0.111) |
| | Labor Productivity | 0.163*** (0.022) | 0.258*** (0.019) | 0.236*** (0.018) |
| | Open-Source Development Intensity (*OSDI*) | 0.014*** (0.005) | 0.012*** (0.004) | 0.016*** (0.004) |
| | Year Fixed Effect | Yes | Yes | Yes |
| | Control Variables† | Yes | Yes | Yes |
| Number of Firm-years | | 747 | 747 | 747 |
| Number of Firms | | 197 | 197 | 197 |
| Pseudo Log-Likelihood | | -181.565 | 29.611 | 215.286 |
| Akaike Information Criterion (AIC) | | 433.129 | 10.779 | -360.571 |
| Bayesian Information Criterion (BIC) | | 594.692 | 172.341 | -199.009 |

*Note:* Standard errors are in parentheses; *** $p<.01$, ** $p<.05$, * $p<.1$; Control variables included *R&D*, *Capital*, *Employees*, *Large Firm*, *Market Share* and *Industry Concentration*, *Total Forks*, *Total Stars*, *Total Issues raised by firm per year*.

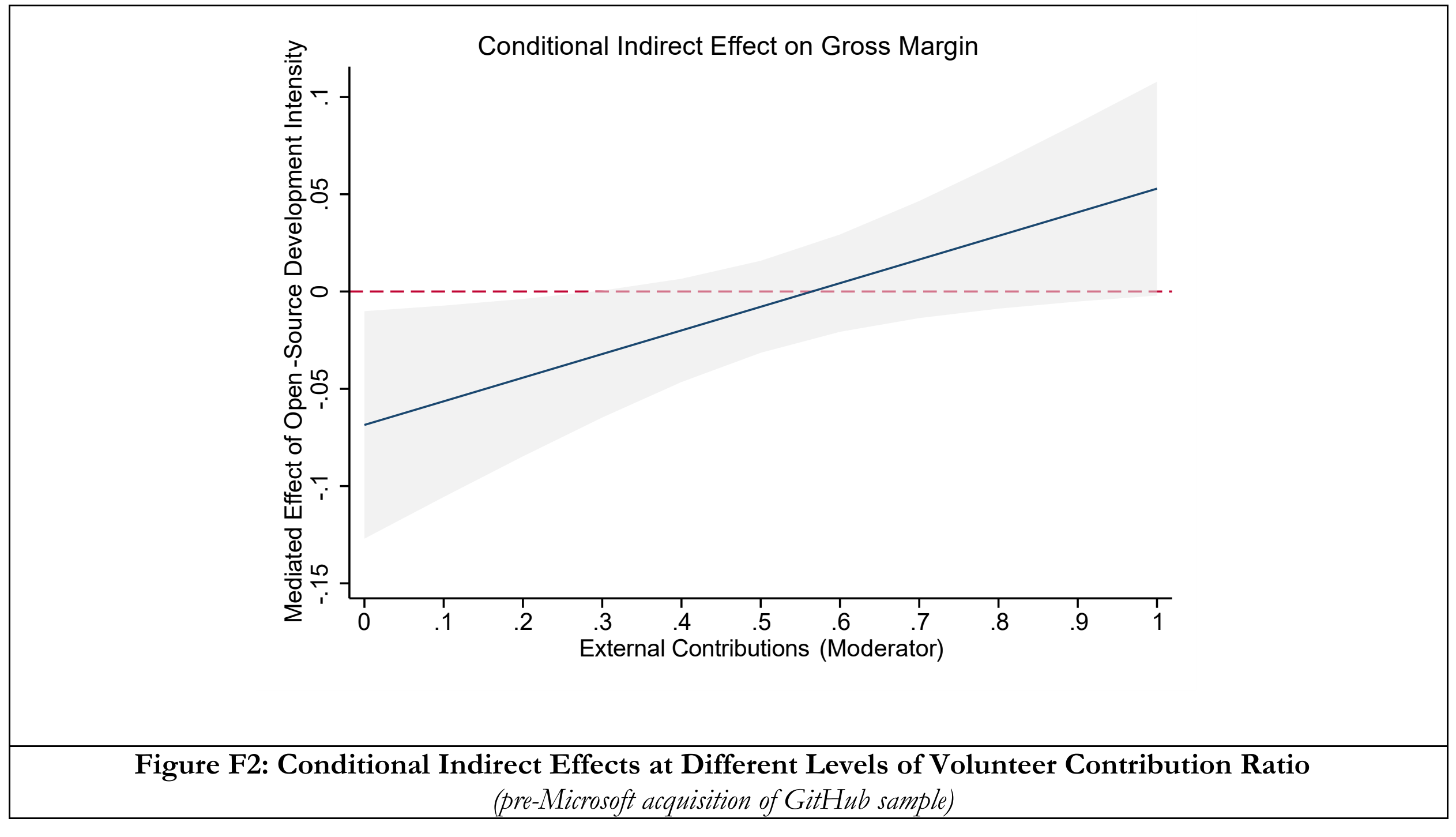

**Figure F2: Conditional Indirect Effects at Different Levels of Volunteer Contribution Ratio**
*(pre-Microsoft acquisition of GitHub sample)*

**Table F2: Open-Source Development Intensity, Labor Productivity, and Profitability**
*Moderated Mediation Analysis using VCR (Full sample, Winsorized)*

| DV | Predictor | Model 1 (*Winsorized at 1%*) | Model 2 (*Winsorized at 5%*) |
|---|---|---|---|
| Labor Productivity | Constant | 2.960***(.147) | 3.216***(.140) |
| | Open-Source Development Intensity (OSDI) | .012**(.006) | .011** (.005) |
| | Volunteer Contribution Ratio (VCR) | -.013 (.053) | -.027 (.051) |
| | OSDI X VCR | .071***(.024) | .090***(.023) |
| | Control Variables† | Yes | Yes |
| | Year Fixed Effect | Yes | Yes |
| Gross Margin | Constant | -.788*** (.076) | -.864*** (.074) |
| | Labor Productivity | .268***(.011) | .248***(.011) |
| | Open-Source Development Intensity (OSDI) | .012***(.002) | .013***(.002) |
| | Year Fixed Effect | Yes | Yes |
| | Control Variables† | Yes | Yes |
| Number of Firm-years | | 1,710 | 1,710 |
| Number of Firms | | 231 | 231 |
| Pseudo Log-Likelihood | | 590.708 | 764.586 |
| Akaike Information Criterion (AIC) | | -1,081.416 | -1,429.172 |
| Bayesian Information Criterion (BIC) | | -809.203 | -1,156.960 |

*Note:* Standard errors are in parentheses; *** $p<.01$, ** $p<.05$, * $p<.1$; Control variables included *R&D*, *Capital, Employees*, *Large Firm*, *Market Share* and *Industry Concentration, Total Forks, Total Stars, Total Issues raised by firm per year.*

**Alternative measure of external contribution: Volunteer Pull Request Ratio**

As a robustness check, we replaced the *Volunteer Contribution Ratio (VCR)* with the *Volunteer Pull Request Ratio (VPRR)*, defined as the share of pull requests submitted by non-firm contributors[1]. Pull-request measures are narrower than unique contributor measures, as they emphasize intensive, code-centric activity and undercount broader volunteer participation such as issue reporting, review comments, documentation, and sporadic micro-contributions (Young et al. 2021; Shen et al. 2025). Prior research shows that many productivity and learning benefits of open-source collaboration stem from broad participation by many contributors rather than from contribution volume alone, which is more fully captured by unique contributor measures (Young et al. 2021). Nevertheless, pull requests represent the primary governance mechanism through which external developers contribute code in firm-managed open-source projects and thus provide a well-established proxy for boundary-spanning technical input (Padhye et al. 2014; Tsay et al. 2014). Empirical studies further link external pull requests to higher productivity, faster issue resolution, and improved software quality when firms retain internal reviewers and integrators (Tsay et al. 2014; Shen et al. 2025).

Figure F3 shows the overall *Volunteer Pull Request Ratio (VPRR)* density in our sample. The *VPRR* has a mean of 0.49 and a median of 0.50, suggesting that roughly half of pull requests are typically volunteer-initiated. The middle 50% of observations lie between 0.33 (25th percentile) and 0.62 (75th percentile), indicating moderate variability, with many projects reaching the upper bound of 1.0. The distribution is nearly symmetric (skewness ≈ 0.06).

[1] Pull requests are not used exclusively by external contributors. Prior research documents that core and project-affiliated developers, including firm employees, also regularly submit pull requests, particularly in repositories that enforce pull-based workflows and mandatory code review. Consequently, pull-request activity reflects a mixture of internal and external development effort (Gousios et al. 2014; Tsay et al. 2014).

We estimated the moderated mediation model using *VPRR* as a moderator Winsorizing at the 1% and 5% thresholds to mitigate the influence of extreme outliers. The results of these tests are presented in Table F3 – Full sample (Model 1), Winsorized at 1% (Model 2), and Winsorized at 5% (Model 3). Across these various specifications, the results remain consistent. *VPRR* moderates the effect of *OSDI* on *Gross Margin* mediated by *Labor Productivity*. Table F4 reports the bootstrapped conditional indirect effects at different levels of *VPRR*. The indirect effect of *OSDI* on *Gross Margin* via *Labor Productivity* is significant at mean (0.003, p<0.05) and high level (+1 SD: 0.005, p<0.01) but not at low level (−1 SD: indirect effect=0.001, p>.1) of external pull request contributions. The Index of Moderated Mediation is significant (ω=0.013, 95% CI [0.002, 0.024]), confirming that the strength of the mediation is statistically dependent on the level of external volunteer pull request contributions, further supporting our main results in the main-text Table 4. The conditional indirect mediated effect of OSDI on Gross Margin is visualized in Figure F4 for the full sample.

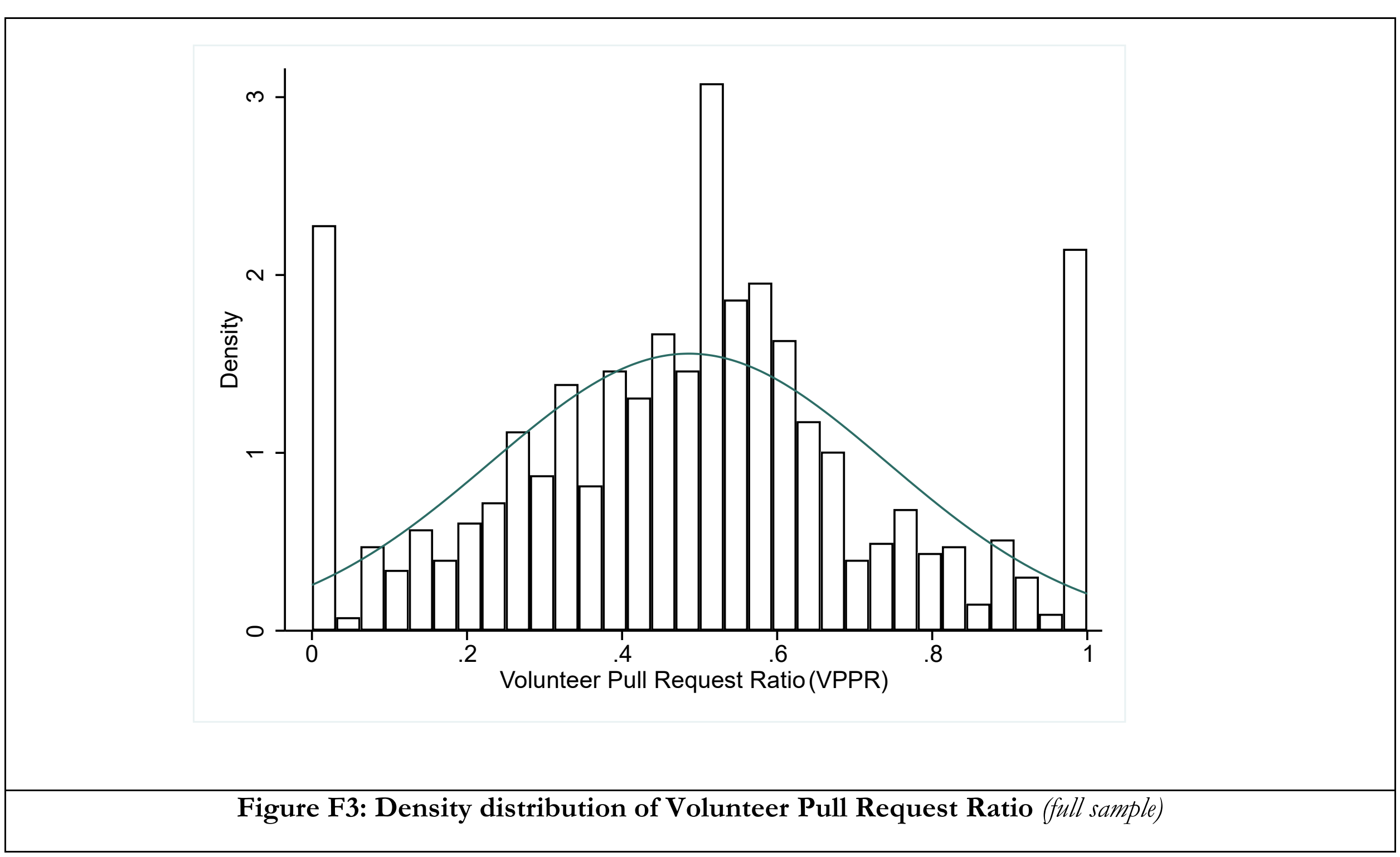

**Figure F3: Density distribution of Volunteer Pull Request Ratio** *(full sample)*

**Table F3: Open-Source Development Intensity, Labor Productivity, and Profitability**
*Moderated Mediation Analysis using VPRR (Full sample)*

| DV | Predictor | Model 1 | Model 2 (*Winsorized at 1%*) | Model 3 (*Winsorized at 5%*) |
|---|---|---|---|---|
| Labor Productivity | Constant | 2.429*** (0.156) | 2.771*** (0.150) | 3.058*** (0.147) |
| | Open-Source Development Intensity (*OSDI*) | 0.016*** (0.006) | 0.017*** (0.006) | 0.015*** (0.006) |
| | Volunteer Pull Request Ratio (*VPRR*) | -0.041 (0.083) | -0.057 (0.080) | -0.071 (0.078) |
| | *OSDI* X *VPRR* | 0.067** (0.030) | 0.076*** (0.029) | 0.084*** (0.028) |
| | Control Variables† | Yes | Yes | Yes |

|  | Year Fixed Effect | Yes | Yes | Yes |
|---|---|---|---|---|
| *Gross* Margin | Constant | -0.294*** (0.090) | -0.792*** (0.076) | -0.860*** (0.068) |
|  | Labor Productivity | 0.189*** (0.013) | 0.259*** (0.012) | 0.236*** (0.010) |
|  | Open-Source Development Intensity (*OSDI*) | 0.009*** (0.003) | 0.010*** (0.003) | 0.012*** (0.002) |
|  | Year Fixed Effect | Yes | Yes | Yes |
|  | Control Variables† | Yes | Yes | Yes |
| Number of Firm-years |  | 1,571 | 1,571 | 1,571 |
| Number of Firms |  | 222 | 222 | 222 |
| Pseudo Log-Likelihood |  | 222.133 | 583.375 | 835.906 |
| Akaike Information Criterion (AIC) |  | -350.265 | -1,072.749 | -1,577.812 |
| Bayesian Information Criterion (BIC) |  | -98.370 | -820.854 | -1,325.917 |

*Note:* Standard errors are in parentheses*; *** p<.01, ** p<.05, * p<.1;* Control variables included *R&D*, *Capital, Employees*, *Large Firm*, *Market Share* and *Industry Concentration, Total Forks, Total Stars, Total Issues raised by firm per year. For some firm-years the volunteer pull request ratio was not available where total external and internal pull requests for the year were zero.*

**Table F4: Bootstrapped Conditional Indirect Effects at Different Levels of Volunteer Pull Request Ratio**

| Levels of VPRR | Indirect Effect | Boot SE |
|---|---|---|
| Low (-1 SD) | 0.001 | 0.002 |
| Mean (0) | 0.003** | 0.001 |
| High (+1 SD) | 0.005*** | 0.002 |
| Index of Moderated Mediation | 0.013** | 0.006 |

*Note: *** p<.01, ** p<.05, * p<.1*

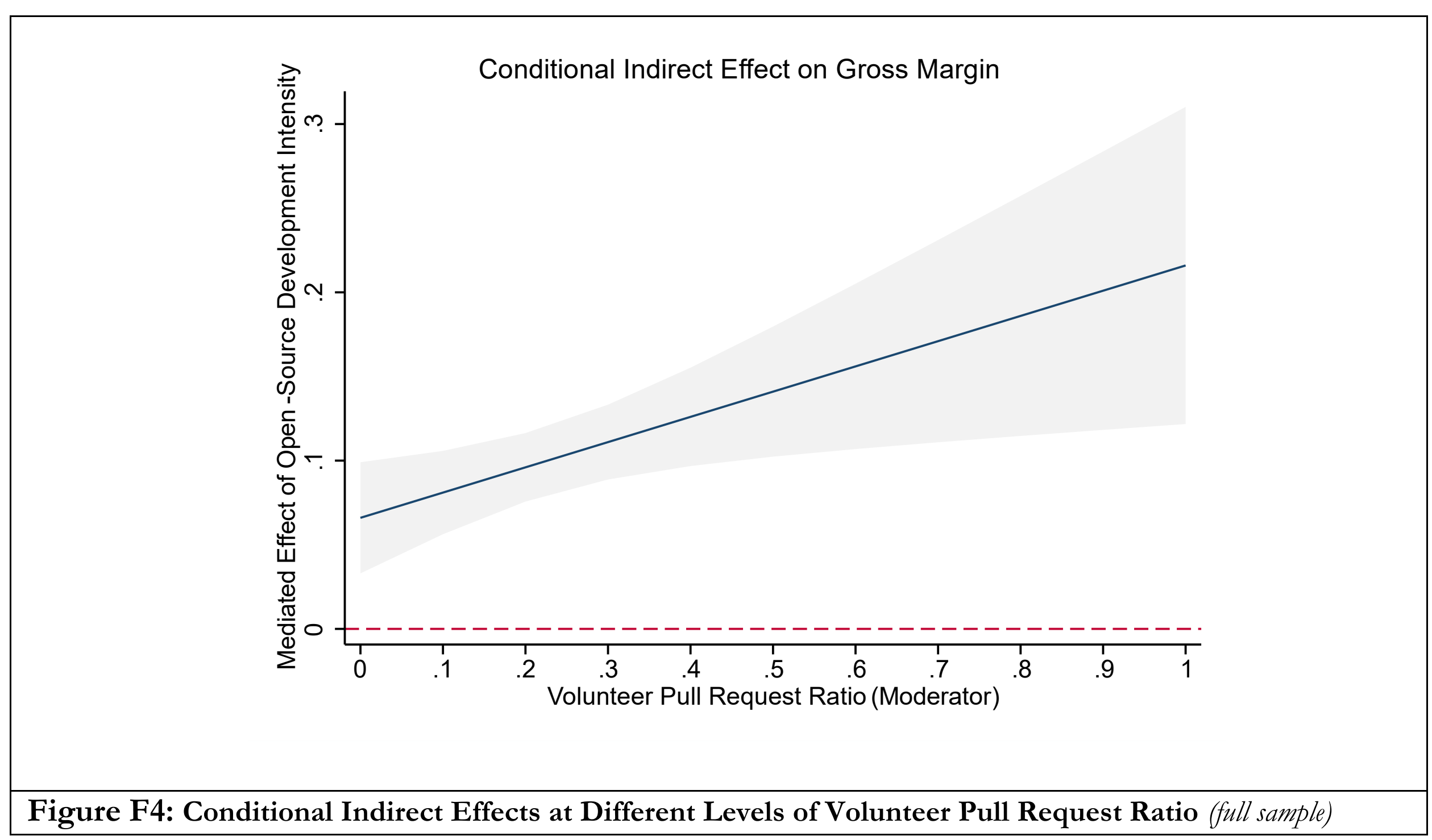

**Figure F4: Conditional Indirect Effects at Different Levels of Volunteer Pull Request Ratio** *(full sample)*